%% file: MDHybrid4YSpectrumELS_arXiv.tex
\documentclass[5p]{elsarticle}
\usepackage{mathtools}
\usepackage{type1cm}
\usepackage{geometry}
\usepackage{color}
\usepackage{latexsym}
\usepackage{setspace}
\usepackage{hyperref}
\usepackage[all]{hypcap}
\usepackage{graphicx}
\usepackage{float}
\usepackage{subfig}
\usepackage{multirow}
\usepackage{array}
\usepackage{rotating}
\usepackage{etoolbox}

\begin{document}
\begin{frontmatter}
\title{The Hybrid Energy Spectrum of Telescope Array's Middle Drum Detector and Surface Array}
\include{TAauthor_list-20140620_formatted}

\end{frontmatter}

\newcommand{\Deg}{^{\circ}} 
\newcommand{\xmax}{$X_{max}$}
\newcommand{\gcm}{g/cm$^2$}
\doublespacing
\include{abstract}

\section{Introduction}
The Telescope Array (TA) experiment is located near Delta, Utah, about 250~km southwest of Salt Lake City. It is a hybrid experiment that incorporates two of the main types of cosmic ray detectors (fluorescence telescopes and a scintillation counter array) for studying Ultra High Energy Cosmic Rays (UHECR).

\begin{figure}[tbp]
  \centerline{
   \includegraphics[width=1\linewidth]{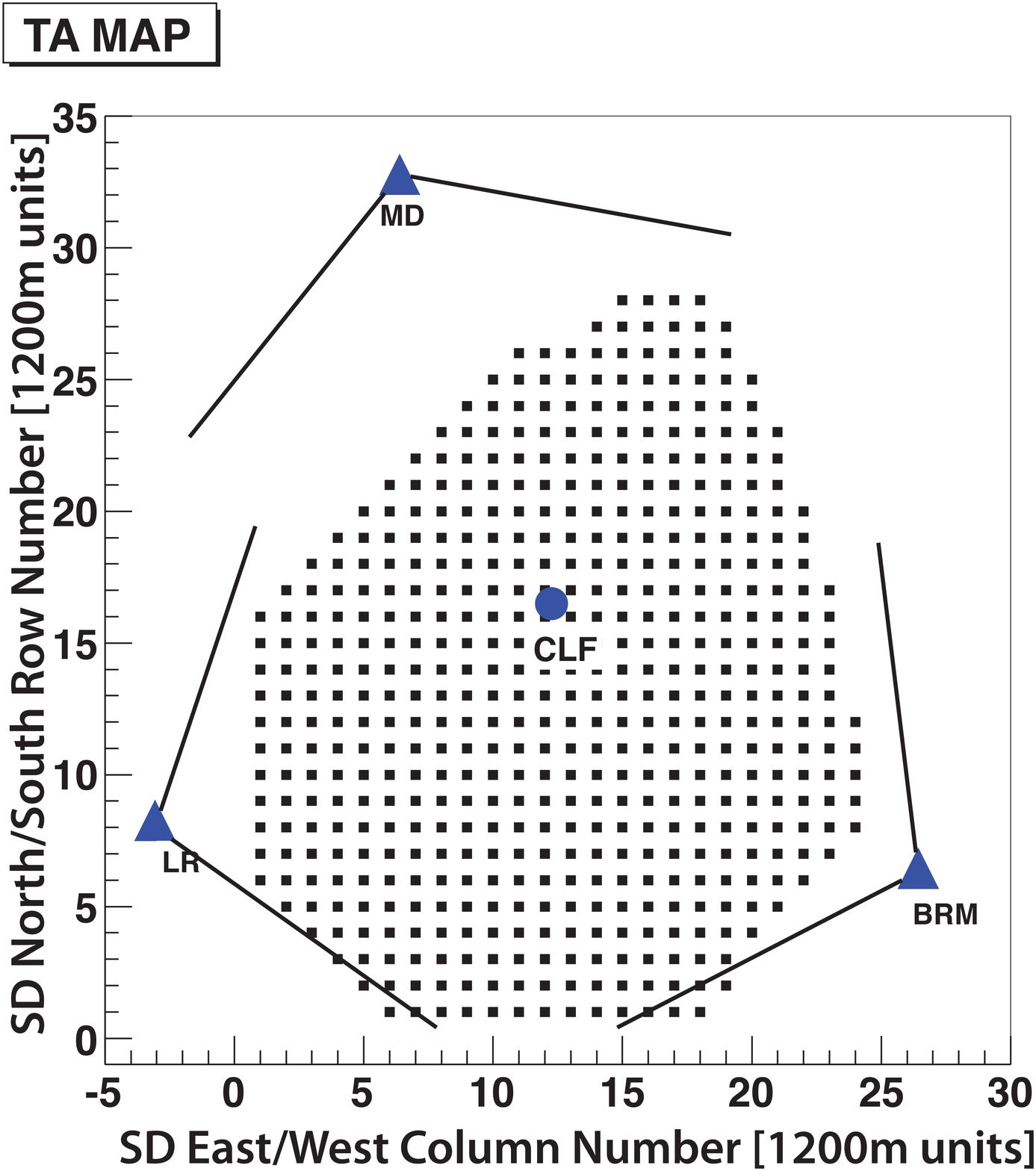}
   }
   \caption[The layout of the Telescope Array experiment.]{The layout of the Telescope Array experiment. The filled black squares indicate the locations of the 507 scintillation counters that comprise the Surface Detector (SD) array. The triangles mark the three fluorescence detector sites at the periphery of the SD array. The solid black lines indicate the field of view for each of the fluorescence detector sites. The Central Laser Facility (CLF), shown by the circle, is placed equidistant from the three fluorescence detector sites to provide atmospheric monitoring and cross-calibration.} 
  \label{fig:map}
\end{figure}

Figure \ref{fig:map} shows the distribution of the 507 scintillation counters that comprise the TA scintillator Surface Detector (SD) array. The locations of the counters, shown by the filled black squares, are laid out on a 1.2~km~square grid. The SD counters sample the laterally-distributed remnants of the air showers at ground level ($\sim$4600~ft above sea level). The SD array is operational 24 hours a day. It rarely has more than a few detectors down at any given time, and often operates with all of them. Taking into account the data acquisition system efficiency, it has a duty cycle of $>$95\%. The detection efficiency of air showers with the SD array rises quickly above $\sim$10$^{18}$~eV and it becomes fully efficient above $\sim$10$^{19}$~eV. The aperture for the highest energy cosmic rays is about 1500~km$^2$ steradians. 

The three Fluorescence Detector (FD) sites, indicated by the triangles in Figure \ref{fig:map}, are located at the periphery of the SD array and view the sky over the array. The two southern sites each consist of 12 new telescopes built for the TA experiment. The northernmost FD site, located at Middle Drum (MD), was constructed with 14 refurbished telescopes from the HiRes-1 site of the previous High-Resolution Fly's Eye (HiRes) experiment. The re-use of these telescopes provides a direct connection between TA and HiRes: the energy scale of the HiRes experiment can be directly transferred to TA. 

In this paper, we introduce the MD hybrid reconstruction method and then compare the resulting spectrum to the measurement results achieved by the MD telescope station and the SD array acting alone. By using the SD and telescope detector in hybrid mode, the geometry reconstruction of the showers is improved significantly, as is shown in section \ref{sec:Resolutions}. A more accurate reconstruction of the geometry leads to a more accurate energy measurement of the primary particle. An initial comparison between this MD hybrid analysis and the MD monocular analysis has been shown in \cite{Rodriguez2012}, along with a detailed comparison between the MD monocular analysis and the HiRes experiment. Here, we intend to take these comparisons a step further by comparing the MD hybrid spectrum to the MD mono spectrum as well as the SD array, linking all parts of the TA measurements to those of the HiRes experiment.


\section{Surface Detectors}
The 507 scintillation counters in the SD array are arranged on a 1.2~km square grid and each have an active area of 3~m$^2$. The spacing and active area were optimized to provide $\sim$100\% detection efficiency for events with energy, E$\geq 10^{19}$~eV. Each detector is composed of two layers of 1.2~cm thick extruded scintillator with grooves in it \cite{Abu-Zayyad2012a, ICRR2006}. Wavelength shifting optical fibers run through the grooves to collect the light generated when particles pass through the scintillator and both ends of the optical fibers run to one of two PMTs in the SD, one PMT per scintillator layer \cite{Abu-Zayyad2012a, ICRR2006}. Each layer of scintillator with optical fibers is wrapped in Tyvek sheeting to help ensure optimum light capture. The average signal from single cosmic ray muons, or a Minimum Ionizing Particle (MIP) is used to calibrate the signal from an event.

\begin{figure}[tbp]
  \centerline{
   \includegraphics[width=0.8\linewidth]{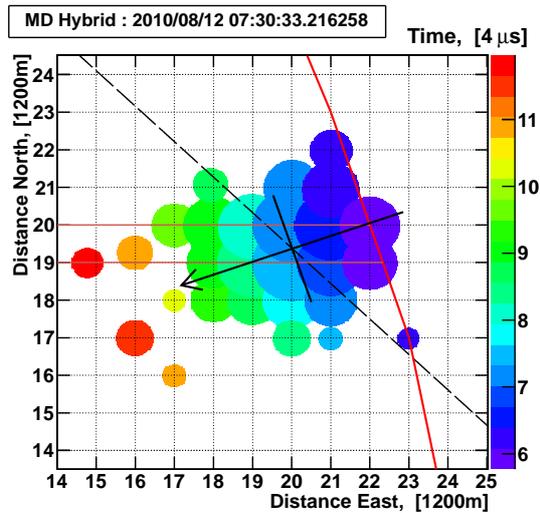}
   }
   \caption[An event display for a typical Surface Detector (SD) event.]{An event display for a typical Surface Detector (SD) event. SD counters are located nearly at the intersection points of the grid. For each detector viewing the event, the circle size is proportional to the number of incident particles on that detector, and circle color represents the trigger timing of each detector. The arrow represents the reconstructed direction of the shower, and the point where the arrow crosses the solid black line represents the reconstructed shower core position on the ground. The red line represents the SD array boundary. The black dashed line represents the line of sight to the core of the shower from the Middle Drum Detector.} 
  \label{fig:SD_event_display}
\end{figure}

The signals from each of the PMTs pass through a shaping circuit and are digitized by a Flash Analog to Digital Converter (FADC), operating at 50~MHz. While the FADC digitizes the analog input from the PMTs, those pulses which exceed 0.3~MIPs in integrated area are stored in memory, along with the time of the pulse (registered via a GPS clock) \cite{Abu-Zayyad2012a, Ivanov2012}. The SD array is divided into three sub-arrays with one wireless control/communications tower over each sub-array. Trigger computers at the communication towers poll each SD counter in their sub-array at one second intervals. The time of pulses greater than 3~MIPs are reported to the towers and this information is used to form an event trigger. An event trigger occurs when three adjacent SDs see a signal greater than 3~MIPs within an 8~$\mu$s window. When an event trigger does occur, the signals from all of the detectors in the array with signal greater than 0.3~MIPs within a coincidence window of $\pm$32~$\mu$s are then transferred from the individual counters first to the tower PC and finally to the central data acquisition system in the city to the east of the TA site via the wireless network \cite{Abu-Zayyad2012a, Ivanov2012}. Figure \ref{fig:SD_event_display} shows an event display of a typical SD event.

\section{Middle Drum Detector}
The MD detector consists of 14 telescopes and is located $\sim$10~km from the nearest SD at the northern end of the array. It is about 21~km northwest of the Central Laser Facility. Each of the 14 telescopes consists of a 5.1~m$^2$ spherical mirror which images the luminous air shower onto a camera comprised of a cluster of 256 PMTs  \cite{Rodriguez2011}.

Each telescope mirror is composed of four glass mirror segments arranged in a cloverleaf shape. The segments are individually adjustable, and have been aligned to focus light onto the camera at their common focal plane. Due to the obscuration of the cluster box and stand directly in front of the mirror, the total effective collection area of the mirror is 3.72~m$^2$. Seven of the 14 telescopes view $3\Deg - 17\Deg$ in elevation, while the remainder view $17 \Deg - 31 \Deg$. In azimuth, all 14 mirrors used in conjunction can see $112 \Deg$ between southwest and southeast.

\begin{figure}[tbp] 
\centerline{
\includegraphics[width=0.8\linewidth]{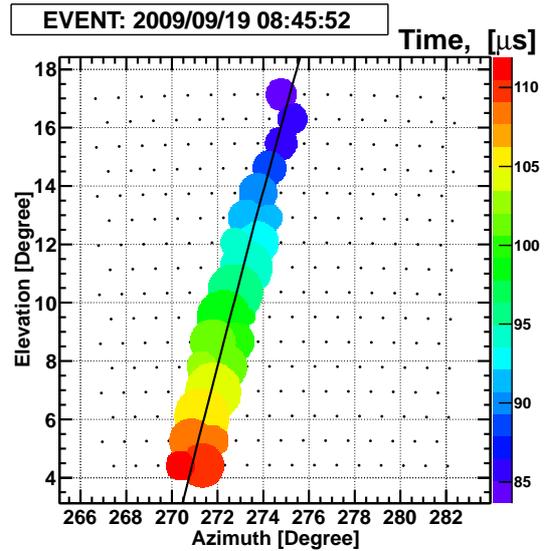}
}
\caption[Middle Drum event display for event 2009/09/19 08:45:52: it shows the Shower Detector Plane (SDP) fit to the PMTs.]{Middle Drum event display for event 2009/09/19 08:45:52. For each PMT, the size of the circle is proportional to the amount of light collected by the PMT, while the color of the circle represents the timing with respect to the other tubes. The black line represents the fit to the projection of the Shower Detector Plane (SDP).} 
\label{fig:MDDisp}
\end{figure}

The fluorescence light collected by the mirror passes through a UV band-pass filter before reaching the PMTs in order to remove most starshine and man made light and thus improve the signal to noise ratio. Within the camera cluster box, are 256 hexagonally close-packed PMTs. Each PMT is optimized to collect UV light and is provided with its own high voltage setting to provide uniform gain. Figure \ref{fig:MDDisp} shows an event display from the MD detector.

Each PMT is individually monitored and the threshold (1240-2500~mV) is continuously modified to keep the tube trigger rate, or ``count rate" at 200~Hz. A single tube trigger is saved for 25~$\mu$s. A ``subcluster" (a 4x4 cluster of 16 PMTs within one camera) trigger occurs when three tubes trigger within a 25~$\mu$s window, and two of them are adjacent. When the conditions are met, the subcluster trigger is transmitted to a ``mirror trigger" board. When two subclusters trigger within a 25~$\mu$s window, a ``telescope" level trigger occurs \cite{Rodriguez2011}. All the PMT signals are converted to a digital signal through a 12-bit Analog to Digital Converter (ADC) \cite{Allen2012}.

\section {Middle Drum Hybrid Event Reconstruction}
The MD hybrid analysis takes advantage of existing programs used to reconstruct events in monocular mode by both the SDs and the FDs. After the initial reconstruction steps are done separately, the events are combined for a hybrid analysis.

\subsection{SD Reconstruction}
The raw data from the SD array contains trigger and waveform information from particles passing through the scintillator and producing light that is detected by the PMTs. The SD reconstruction determines the geometry and energy of the events from these signals. The FADC traces are scanned to find the time of the signal. It is then calibrated using the 1 MIP from detected cosmic ray muons. The calibrated information from the triggered events is used to fit the geometry of the shower. First, the counters with signals from the actual event are identified. This is done by only including counters which are considered contiguous in both space and time. Counters that are within $\sqrt2 \, \times$ the counter spacing are considered contiguous in space, thus including counters on the diagonal. Two counters with a time difference (divided by the speed of light) less than or equal to their spacing are considered contiguous in time. Counters that don't fit this pattern recognition criteria are removed as electronic noise or random muons. The shower track vector indicating the geometry of the shower is then found using the trigger times of each SD in the event.

In the final reconstruction step for the SD events, the triggered counters are fit to a Lateral Distribution Function (LDF). The SD array is a direct derivative of the basic design of the Akeno Giant Air Shower Array (AGASA) experiment, though it is optimized to detect events with higher energies by increasing the spacing and the detector size. Therefore, it makes sense that the SD reconstruction programs use the same LDF that was used by the AGASA experiment \cite{Ivanov2012, Takeda2002}. This was done so that a good comparison could be made between the TA surface array and the AGASA experiment. Such a comparison has been done \cite{Ivanov2012}. Using the result of the LDF geometry fit, the density of particles at a lateral distance perpendicular to the shower core can be extrapolated at any point. Studies have shown that the optimum parameter for determining the energy of an air shower using a ground array is the signal at a fixed distance from the shower core. That specific distance is dependent primarily on array geometry, and has little dependence on shower geometry or the lateral distribution function that is used \cite{Newton2007}. The distance $\sim$800~m from the shower core has been determined to be a stable indicator of shower energy for this size detector (3~m$^2$) and counter separation (1.2~km) \cite{Ivanov2012}. The density of particles at this point is called S800. Once S800 is found, an energy table created from the Monte Carlo (MC) (described in section \ref{sec:Simulation}) is used to determine the energy. The table is generated by matching the original thrown energy of the Monte Carlo showers to the final reconstructed values, including S800 and zenith angle. In this method, many simulated showers with different energies and geometries are generated to find the one which gives signals in the detector which most closely resemble the actual data event. 

\subsection{MD Reconstruction}
The FD reconstruction for MD begins by matching the triggered events from individual telescopes using GPS time-stamps. The data from the telescopes are then compared, and telescope triggers that occur within 100~$\mu$s of each other are combined into a single site event. The reconstruction program then determines the probability that a given event was triggered by noise using a Rayleigh filter. Each pair of neighboring tubes is examined and a unit vector is drawn from the earlier tube to the later one. A Rayleigh vector describes the sum of all such segments for a given event. If the event is due to noise, the length of the Rayleigh vector will be short, while for a real cosmic ray event it will be long. Using the Rayleigh vector, a probability that the event was triggered randomly is calculated. Each event that has a probability of 1\% or less of having been generated by noise is saved for further analysis. Using the pointing directions of the PMTs, the Shower Detector Plane (SDP) is calculated for each of the saved events. The SDP is treated as a line source and is fit using $\chi^2$ minimization for Equation \ref{eq:SDPchi2}.
\begin{equation}
\chi^2 = \sum_i \frac{(\mathbf{\hat{n}} \cdot \mathbf{\hat{n_{\emph{i}}}})^2 \cdot w_i}{\sigma_i^2}
\label{eq:SDPchi2}
\end{equation}

\noindent In this equation, $\mathbf{\hat{n}}$ represents the SDP normal vector, and $\mathbf{\hat{n_i}}$ is the viewing direction of triggered tube $i$. The number of photoelectrons seen by tube $i$ is $w_i$. For each tube, $\sigma_i$, or the angular uncertainty, is set to 1$\Deg$ because this is the field of view of an individual PMT and we can not determine where a photon hits on the face of the PMT. Finally, the program looks for groups of events that are similar in time, core location, and amount of light seen, with a goal of removing those events that are from artificial sources. These removed sources would include laser shots from the Central Laser Facility, which are routinely made for atmospheric monitoring.

\subsection{Hybrid Reconstruction}
As described in the previous sections, the SD and MD events are reconstructed separately through the SD and the MD reconstruction programs. In order to combine the two sets of information into one hybrid event set, a time matching program compares the two data sets. The time that the shower core intersects with the ground, or plane in which the SDs lie, is calculated for each set and compared. Events that are within 2~$\mu$s of each other are considered matched. They are combined into a single common hybrid event.

Once a combined set has been created, the events are reprocessed using the information from both detectors. We minimize the $\chi^{2}$ taking into account (1) the Fluorescence Detector timing, (2) the SD timing, and (3) the position of the core of the shower as it hits the ground as determined by the SD, including uncertainties.

The timing of the FDs and SDs is combined by comparing timing with pointing direction. Using Equation \ref{eq:TVSA},  MD PMT trigger times can be related to their pointing direction. The resulting $\chi^{2}$ for minimization is then shown by Equation \ref{eq:FDChi2Tim}. 
\begin{align}
t_{i} &= T_{Rp} + \frac{R_{P}}{c} tan \left( \frac{\pi - \psi - \chi_{i} }{2} \right)
\label{eq:TVSA} \\
\chi^{2}_{MD_{Timing}} &= \sum_{i} \frac{1}{\sigma^{2}_{i}}  \left[ t_{i} - \left( T_{Rp} + \frac{R_{P}}{c} tan \left( \frac{\pi - \psi - \chi_{i} }{2} \right) \right) \right]
\label{eq:FDChi2Tim}
\end{align}

\noindent In both equations, $t_i$ represents the triggered time of tube $i$, and $T_{Rp}$ represents the time of the shower (in microseconds) at the impact parameter ($R_P$), measured in km. The angle of the shower track within the SDP is represented by $\psi$ (in degrees), and $\chi_i$ is the tube viewing angle within the SDP. 

\begin{figure}[tbp] 
\centerline{
\includegraphics[width=\linewidth]{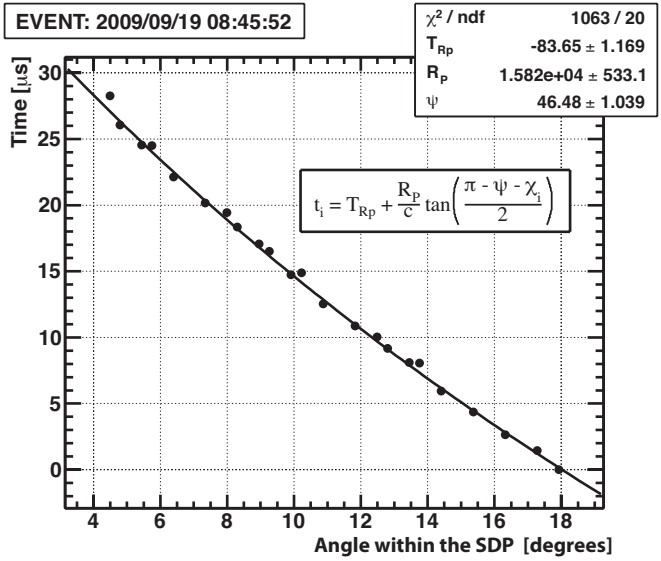}
}
\caption[Timing vs angle plot for event 2009/09/19 08:45:52, observed by the PMTs at the Middle Drum fluorescence detector site: the angle of the observed signal along the SDP is plotted with respect to the time information of the signal.]{Timing vs angle plot for event 2009/09/19 08:45:52, observed by the Middle Drum fluorescence detector site. The angle of the observed signal along the Shower Detector Plane (SDP) is plotted with respect to the time information of the signal. Fitting the curvature provides the timing and impact parameter and, when combined with the SDP, gives the pointing information of the primary cosmic ray.} 
\label{fig:tvsaFDcomp}
\end{figure}

Figure \ref{fig:tvsaFDcomp} shows an example Timing vs. Angle plot of a fluorescence event observed using the MD telescopes. The fit curve is calculated from Equation \ref{eq:TVSA} and $\chi^2$ minimization is used to determine the in-plane angle ($\psi$), impact parameter ($R_{P}$), and time at $R_{P}$ ($T_{Rp}$). 

\begin{figure}[tbp] 
\centerline{
\includegraphics[width=\linewidth]{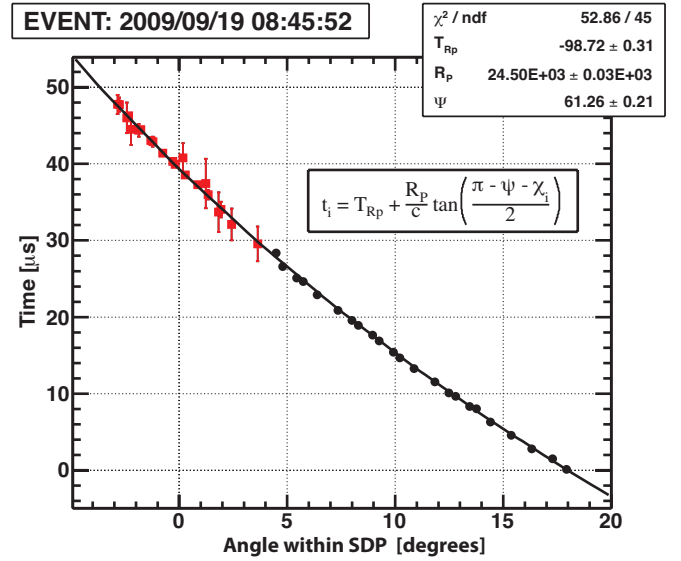}
}
\caption[Timing vs angle plot for event 2009/09/19 08:45:52: it is extended using information from Surface Detectors.]{Timing vs angle plot for event 2009/09/19 08:45:52: it is extended using information from Surface Detectors. Virtual PMTs are created using information from the SD counters (red squares) which have been added to the information from the MD PMTs (black circles). In comparison with Figure \ref{fig:tvsaFDcomp}, the curvature is more obvious, and the $\chi^2$ value is significantly better here, after adding the extra information.} 
\label{fig:tvsaHBcomp}
\end{figure}

Figure \ref{fig:tvsaHBcomp} shows the result of the Hybrid Timing vs. Angle analysis. While Figure \ref{fig:tvsaFDcomp} shows only the MD points, the hybrid plot (Figure \ref{fig:tvsaHBcomp}) has been significantly extended using the timing information from the SDs. Each triggered SD is treated as a virtual PMT located at the MD detector. Equation \ref{eq:SDtimeCalc} shows how the trigger time is adjusted for the SD points.
\begin{equation}
t_{SD} = t_{SD_{Trig}} + \frac{SD_{Dist}}{c}
\label{eq:SDtimeCalc}
\end{equation}

\noindent Here, $t_{SD}$ is the trigger time of a virtual tube at the MD site that represents the position of the counter, while $t_{SD_{Trig}}$ is the actual trigger time of the counter. $SD_{Dist}$ is the distance from MD to the counter, and $c$ is the speed of light. Equation \ref{eq:SDChi2Tim} shows how the SDs are added to the overall $\chi^2$ calculation.
\begin{equation}
\chi^{2}_{SD_{Timing}} = \sum_{i} \frac{1}{\sigma^{2}_{i}}  \left[ t_{i} - \left( T_{Rp} + \frac{R_{P}}{c} tan \left( \frac{\pi - \psi - \chi_{i} }{2} \right) \right) \right]
\label{eq:SDChi2Tim}
\end{equation}

\noindent Note that the equation is the same as Equation \ref{eq:FDChi2Tim}. The difference is that the observed time, $t_i$, is calculated for each SD counter. The signals observed by the SDs arrive later than those measured by the PMTs at the MD detector because the SDs are sampling the shower on the ground, and the light then takes time to get from that point to the telescope. Therefore, as shown in Figure \ref{fig:tvsaHBcomp}, all of the SD points are plotted at later times. Adding the SD counters to the calculation increases the total number of points in the $\chi^2$ minimization and, more importantly, extends the range in time and angle. Note that in comparison with Figure \ref{fig:tvsaFDcomp}, the curvature in Figure \ref{fig:tvsaHBcomp} is more obvious, and the $\chi^2/dof$ is improved. As a result, a more accurate calculation of the geometry is achieved.

The final piece of the $\chi^2$ minimization is the core constraint of the hybrid analysis. Equation \ref{eq:Chi2Core} shows the minimization to determine the x and y coordinates on the ground.
\begin{equation}
\chi^2_{Core} = \sum_1^2 \frac{\|\mathbf{R}_i - \left(\mathbf{R}_{COG}\right)_{i}\|^2}{\sigma^2_{\mathbf{R}_{COG}}}
\label{eq:Chi2Core}
\end{equation}

\noindent Here, $\mathbf{R}_{COG}$ represents the reconstructed core position from the SD Center Of Gravity, $COG$, while $\mathbf{R}_i$ represents the trial parameters. Note that $i=1$ corresponds to the x-coordinate and $i=2$ corresponds to the y-coordinate. The $\sigma_{\mathbf{R}_{COG}}$ is equal to 170~m, the uncertainty determined by the SD Monte Carlo reconstruction \cite{TelescopeArrayCollaboration2014, Ivanov2012}.

The hybrid analysis uses the result of the fit of the SDP normal, $\mathbf{\hat{n}}$, from the MD reconstruction (Equation \ref{eq:SDPchi2}) and varies the parameters $\psi$, $T_{Rp}$, and $R_{P}$ to minimize the full $\chi^2$, including the timing from the SD's, FD's, and the core constraint, simultaneously. This fitting results in the hybrid geometry reconstruction of the UHECR shower. 

The hybrid analysis uses the same energy reconstruction program as the MD monocular processing. It uses an inverse Monte Carlo technique for calculating the shower energy. In order to do this, however, it must first generate a profile of the shower. Using the calculated hybrid geometry, the program converts the viewing angle of each ``good" PMT into a shower depth, in \gcm. 

The Monte Carlo showers for this purpose are parametrically calculated using Poisson statistics rather than thrown and saved. The input parameters for the profile of the calculated shower are taken from the Gaisser-Hillas function, (Equation \ref{eq:GHf}).
\begin{equation}
\label{eq:GHf}
N_{e}\left(x\right)  = N_{max} \times \left[\frac{x-X_{0}}{X_{max}-X_{0}}\right]^{\frac{X_{max}-X_{0}}{\lambda}} exp\left(\frac{X_{max}-x}{\lambda}\right) \, ,
\end{equation}

\noindent The function predicts the number of particles, $N_e$, at a given slant depth, $x$. The values of $X_0$ and $\lambda$ are fixed to 40~\gcm and 70~\gcm, respectively, while, \xmax, representing the depth of the shower maximum, and $N_{max}$, the number of particles at the shower maximum, are allowed to vary. The $\chi^2$ function for the profile is then calculated comparing the number of photoelectrons measured in each PMT to the predicted number calculated for an input shower. The shower with the minimum $\chi^{2}$ corresponds to the Monte Carlo generated shower that best matches the observed shower. The missing energy is estimated by comparing the integrated energy from the visible part of the shower to the original energy of the primary particle in the simulation. The energy of the primary particle from the Monte Carlo shower is then stored as the calculated hybrid energy of the real shower. 

Additional cuts were made on the data using the resolution plots to improve the quality of the reconstruction. Below is a list of quality cuts that were made on the data, based on a study of the simulated showers.

\begin{enumerate}
\item Failmode: Events that failed the profile reconstruction are removed from the set. 
\item Zenith angle $>$ 56$\Deg$: Events with zenith angles greater than 60$\Deg$ cannot be reconstructed reliably, using the SD technique. Therefore, the Monte Carlo for this analysis does not simulate showers with zenith angle greater than 60$\Deg$. Due to overflow, caused by the effect of angular resolution, events close to 60$\Deg$ are also difficult to analyze. 56$\Deg$ is safely distant from 60$\Deg$ for the analysis.
\item Hybrid/SD Core Position (difference $>$1200~m): Since the events are time-matched, it is conceivable that two independent events (one SD event and one MD event) may be combined due to their proximity in time. The core location of the shower at the ground calculated using only the SDs is compared to the position calculated using the hybrid analysis in order to ensure that the MD event and the SD event are the same event, so that only true hybrid events are kept. 
\item Border Cut ($<$100~m): The border cut uses the hybrid core location to determine how close the shower falls to the edge of the SD array. Showers with calculated core locations that fall at, or outside, the border of the array are difficult to reconstruct due to the missing information that may be out of range of the SDs. Therefore, showers with a core that is within 100~m of the border or outside the array are removed.
\item Track Length $<$8.0$\Deg$: Events with shorter track lengths have less information, and therefore provide a less accurate reconstruction.   
\item \xmax \, not ``Bracketed": Events which reconstruct with the depth of the shower maximum, or \xmax\ outside of the field of view of the detector camera (3-31$\Deg$ elevation) are removed. The energy is reconstructed more accurately if \xmax\ is seen.
\end{enumerate}

\section {Simulation} 
\label{sec:Simulation}

An accurate measurement of the hybrid energy spectrum depends upon an understanding of the aperture and exposure of the hybrid detector. The aperture of the detector is dependent upon the layout and efficiency of the detector as well as on the geometry and energy of the shower. Monte Carlo simulations are used to make these calculations.

Simulated events are thrown such that the core of the shower intersects with the ground, or plane in which the SDs lie, within a circle of radius 25~km centered at the Central Laser Facility (CLF), which is at the center of the SD array, equidistant from all three telescope stations. The solid angle, $\Omega_{0}$, is defined by Equation \ref{eq:SolidAngle}. Equation \ref{eq:ApertureThrown} represents the ``thrown" aperture and is defined by the area of the circle multiplied by the solid angle. The calculated aperture for the spectrum is given in Equation \ref{eq:Aperture}.
\begin{align}
\Omega_0 &= 2\pi \int^{\theta_{max}}_0 sin\theta cos\theta d\theta = \pi sin^{2}\theta_{max} 
\label{eq:SolidAngle} \\
A_0\Omega_0 &= \pi^{2} R^{2} sin^{2}\theta_{max} 
\label{eq:ApertureThrown} \\
A\Omega &= A_0\Omega_0\frac{N_{Reconstructed}}{N_{Thrown}}
\label{eq:Aperture}
\end{align}

\noindent Here, $R$ is the radius of the circle (25~km), $\theta_{max}$ is 60$\Deg$ (The maximum zenith angle thrown in the simulated showers), $N_{Reconstructed}$ represents the number of Monte Carlo events that are reconstructed and pass cuts, and $N_{Thrown}$ represents the number of events that were thrown (generated) in the set.

The MC programs simulate both the cosmic ray showers as well as the detector response. The MC showers used for this hybrid analysis were generated using CORSIKA \cite{Heck1998}. At high shower particle energies ($E > 80$~GeV), the QGSJET-II-03 \cite{Ostapchenko2004} hadronic model was used to simulate particle interactions within the shower. At lower energies ($E < 80$~GeV), the FLUKA \cite{Battistoni2006} model was used. The electromagnetic component of the shower was treated using EGS4 \cite{Nelson1985} . 

Over 16,000 dethinned \cite{Stokes2012} proton showers ranging in energy from $10^{16.75}$~eV to $10^{20.55}$~eV with a variety of geometries were created and stored in a shower library \cite{Ivanov2012}. This library was resampled thousands of times using random azimuthal and zenith angles, as well as timing to generate a set of over 150 million simulated events. The set was generated using a piece-wise power law spectrum in a method similar to that used for the HiRes measurement \cite{Abbasi2007}. The following list summarizes the parameters of this main simulated data set.

\begin{itemize}
\item Composition: We assume pure protons and the QGSJET-II-03 hadronic model, which gives good agreement with all geometric variables needed to calculate acceptance. \cite{Abbasi2014}. 
\item Energy Slope, E: $E^{-3.25}$ for $10^{16.75} < E < 10^{18.65}$~eV; 
$E^{-2.81}$ for $10^{18.65}$~eV $\le E < 10^{19.75}$~eV; 
$E^{-5.1}$ for $E \ge 10^{19.75}$~eV. 
This is the piece-wise power law that results from a fit to the HiRes data \cite{Abbasi2007}. 
\item Surface Impact Position: Uniform, random distribution inside a circle of radius 25~km, centered at the CLF (39.296918 N Lat, 112.908733 W Long).
\item Zenith Angle, $\theta$: sin($\theta$)cos($\theta$) distribution in $[0\Deg - 60\Deg]$ range. The sin($\theta)$ represents a spherically isotropic distribution from the sky, while the cos($\theta)$ represents the projection of the distribution on a flat target. 
\item Azimuthal Angle, $\phi$: Flat distribution in $[0\Deg,360\Deg]$ range. 
\end{itemize}

\subsection{Resolutions}
\label{sec:Resolutions}
\begin{figure}[tbp]
\centerline{
\includegraphics[width=0.9\linewidth]{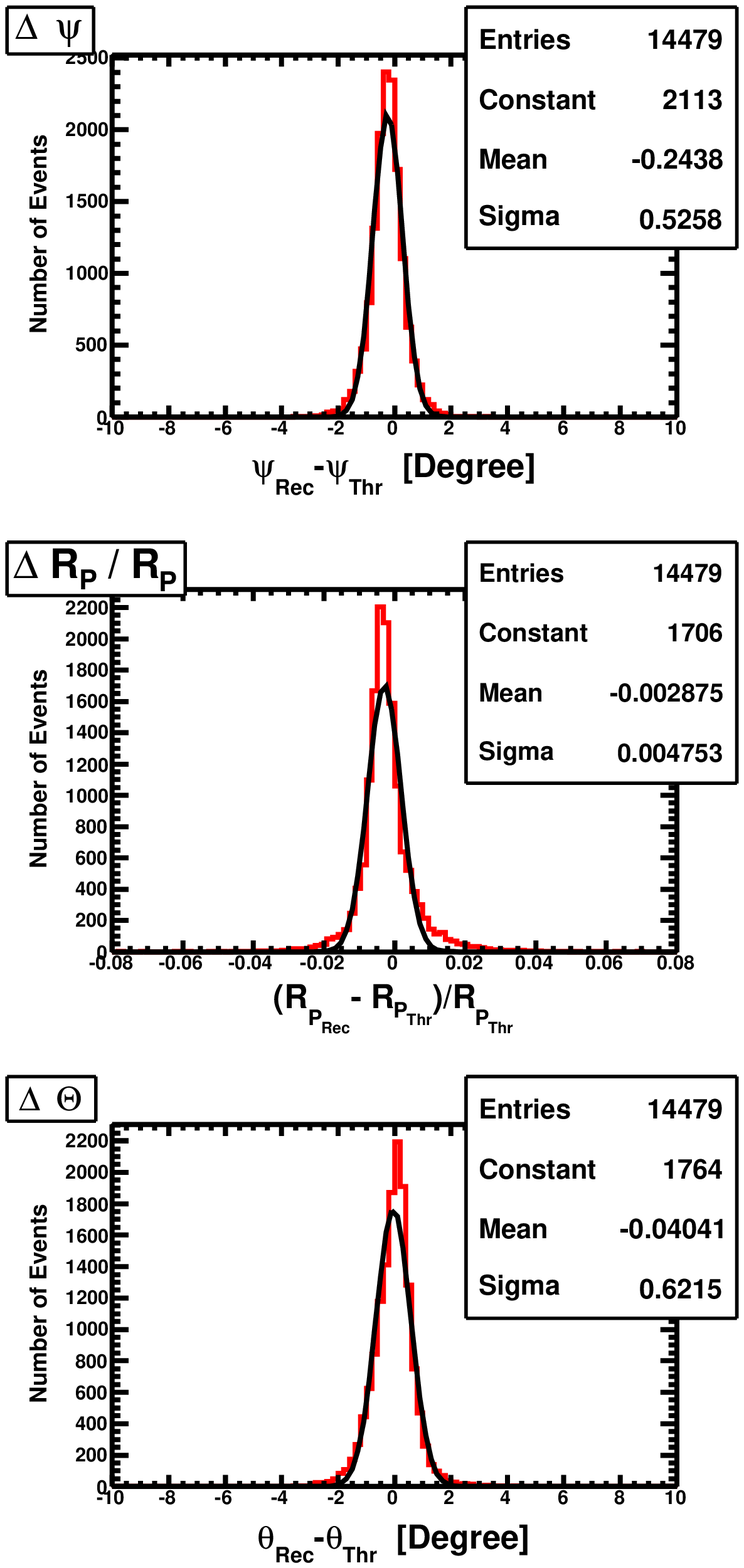}
}
\caption[Resolutions for Middle Drum hybrid geometric reconstructed parameters: shown are the in-plane angle ($\psi$), (top), impact parameter ($R_{P}$), (middle), and zenith angle ($\theta$) (bottom).]{Resolutions for Middle Drum hybrid geometric reconstructed parameters: shown are the in-plane angle ($\psi$), (top), impact parameter ($R_{P}$), (middle), and zenith angle ($\theta$) (bottom). The red histogram shows the difference between the reconstructed and thrown values for each event, or in the case of the impact parameter, the normalized difference. The black line is a gaussian fit to the histogram. Note that the horizontal scale in the hybrid case is different from the monocular reconstruction (shown in the next figure). This reflects the significant improvement in the reconstruction due to the hybrid constraints.}
\label{fig:resGeom}
\end{figure}

\begin{figure}[tbp]
\centerline{
\includegraphics[width=0.9\linewidth]{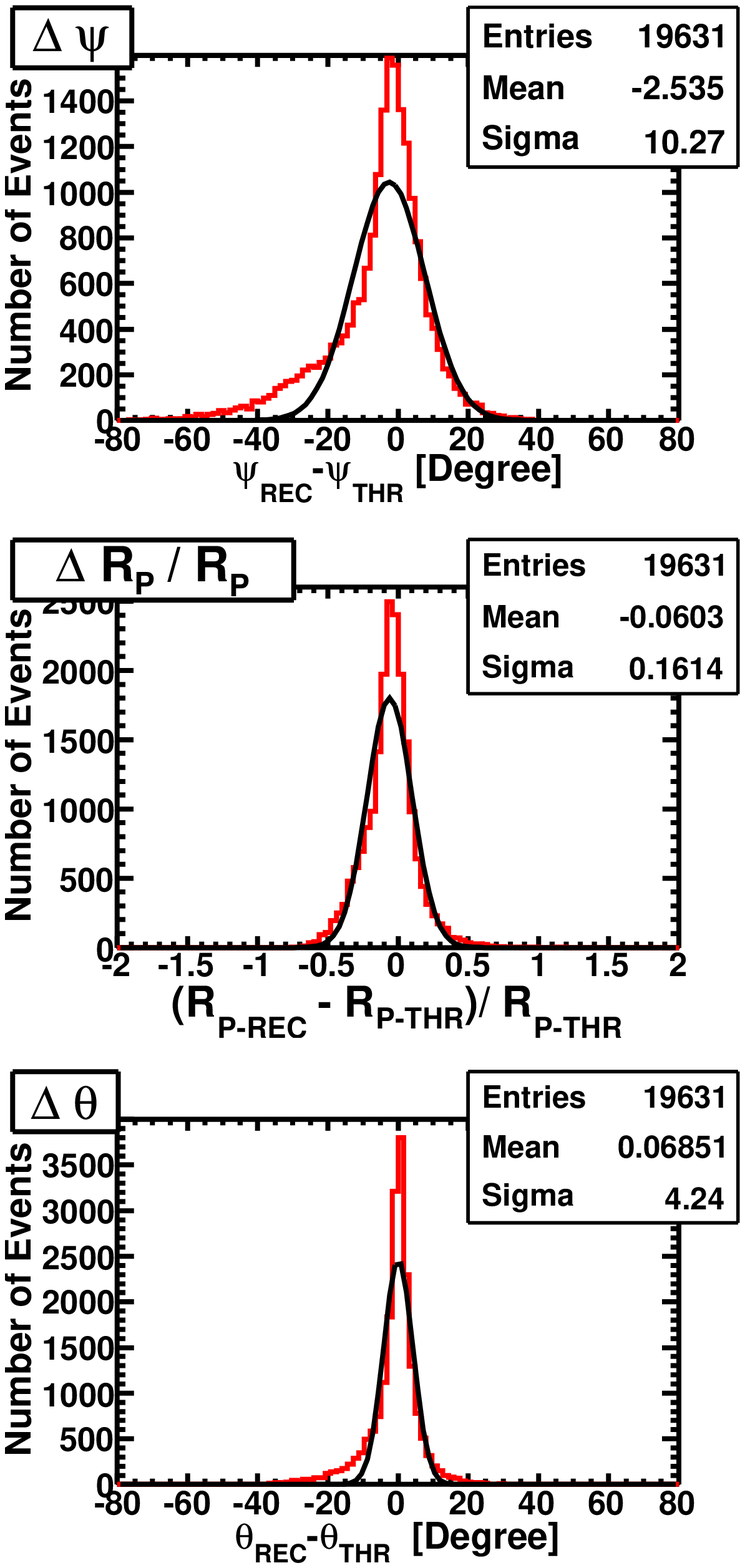}
}
\caption[Resolutions for Middle Drum monocular geometric reconstructed parameters: shown are the in-plane angle ($\psi$), (top), impact parameter ($R_{P}$), (middle), and zenith angle ($\theta$) (bottom).]{Resolutions for Middle Drum monocular geometric parameters: shown are the in-plane angle ($\psi$), (top), impact parameter ($R_{P}$), (middle), and zenith angle ($\theta$) (bottom). The red histogram shows the difference between the reconstructed and thrown values for each event, or in the case of the impact parameter, the normalized difference. The black line is a gaussian fit to the histogram. Note that the horizontal scale in the monocular case is different from the hybrid reconstruction (shown in the previous figure). This reflects the significant improvement in the reconstruction due to the hybrid constraints.}
\label{fig:FDresGeom}
\end{figure}

In the hybrid analysis, both the SD data and MD data are used to constrain the geometrical fit parameters, as detailed in the previous section. In Figure \ref{fig:resGeom}, the reconstructed values of the in-plane angle ($\psi$), impact parameter ($R_P$), and zenith angle ($\theta$) are compared with the MC generated values from the same events. The width of these resolutions from the reconstruction of MC events is used to place an uncertainty on the reconstructed values of the data events. The plots show that the in-plane angle and zenith angle have hybrid resolutions of $\sim$0.5$\Deg$, and the impact parameter has a 0.5\% resolution. Figure \ref{fig:FDresGeom} shows the MD monocular reconstruction resolutions for comparison. The MD hybrid resolutions show significant improvement over the MD monocular reconstruction.

\begin{figure}[tbp]
\centerline{
\includegraphics[width=0.9\linewidth]{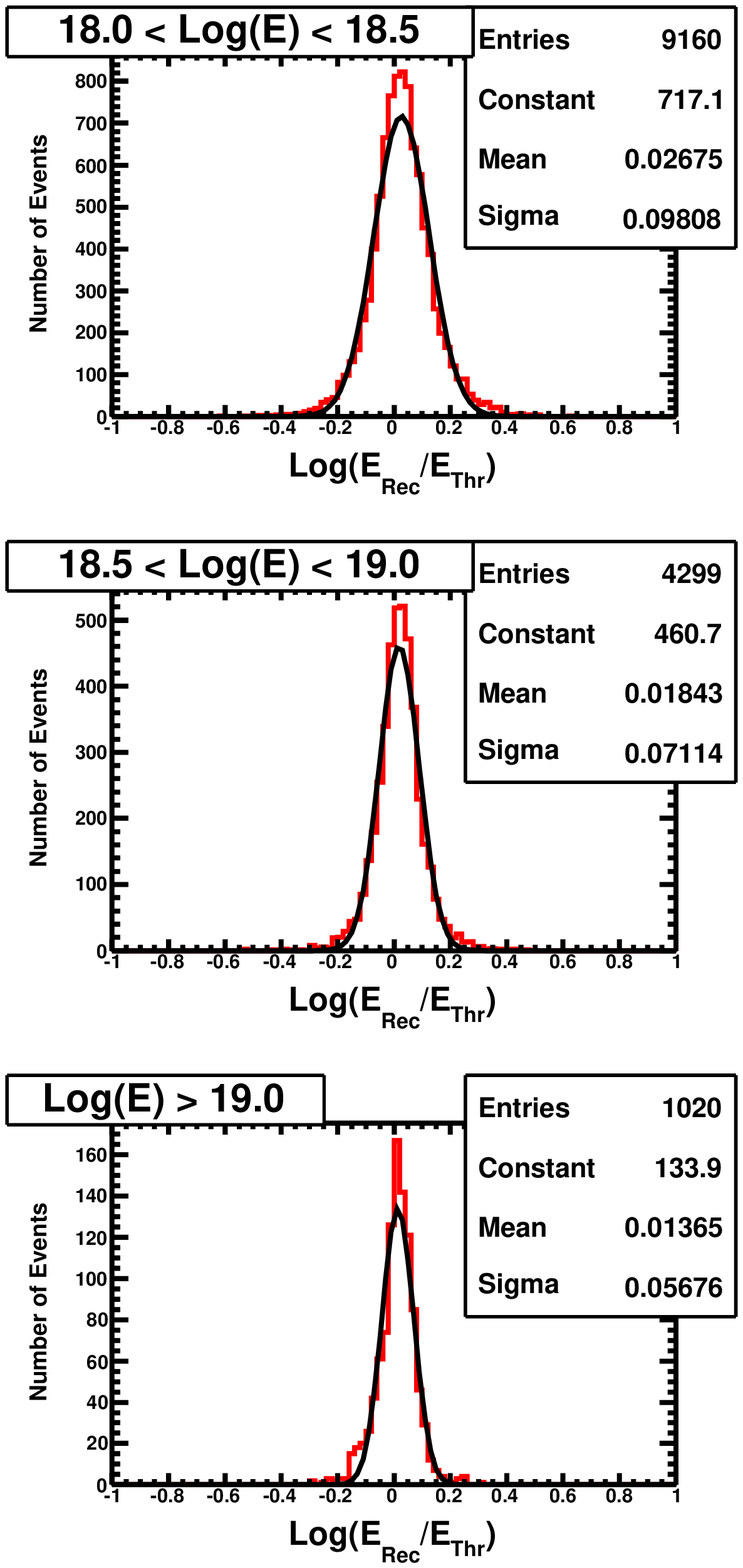}
}
\caption[Resolutions for Middle Drum hybrid reconstructed energy: shown are events broken up into energy ranges: $10^{18}$ $<$ E $<$ $10^{18.5}$~eV (top),  $10^{18.5}$ $<$ E $<$ $10^{19.0}$~eV (middle), and E $>$ $10^{19.0}$~eV (bottom).]{Resolutions for Middle Drum hybrid reconstructed energy: events are shown by energy range: $10^{18}$ $<$ E $<$ $10^{18.5}$~eV (top),  $10^{18.5}$ $<$ E $<$ $10^{19.0}$~eV (middle), and E $>$ $10^{19.0}$~eV (bottom). In each case, the red histogram shows the log of the ratio of the reconstructed and thrown energy for each event. The black line is a gaussian fit to the histogram. The energy resolutions (10\%, 7\%, and 6\%) for the hybrid reconstruction represent more than a factor of two improvement over the monocular reconstruction (34\%, 26\% and 19\%) (shown in the next figure). Note that the horizontal scale is changed in the monocular case.}
\label{fig:resEng}
\end{figure}

\begin{figure}[tbp]
\centerline{
\includegraphics[width=0.9\linewidth]{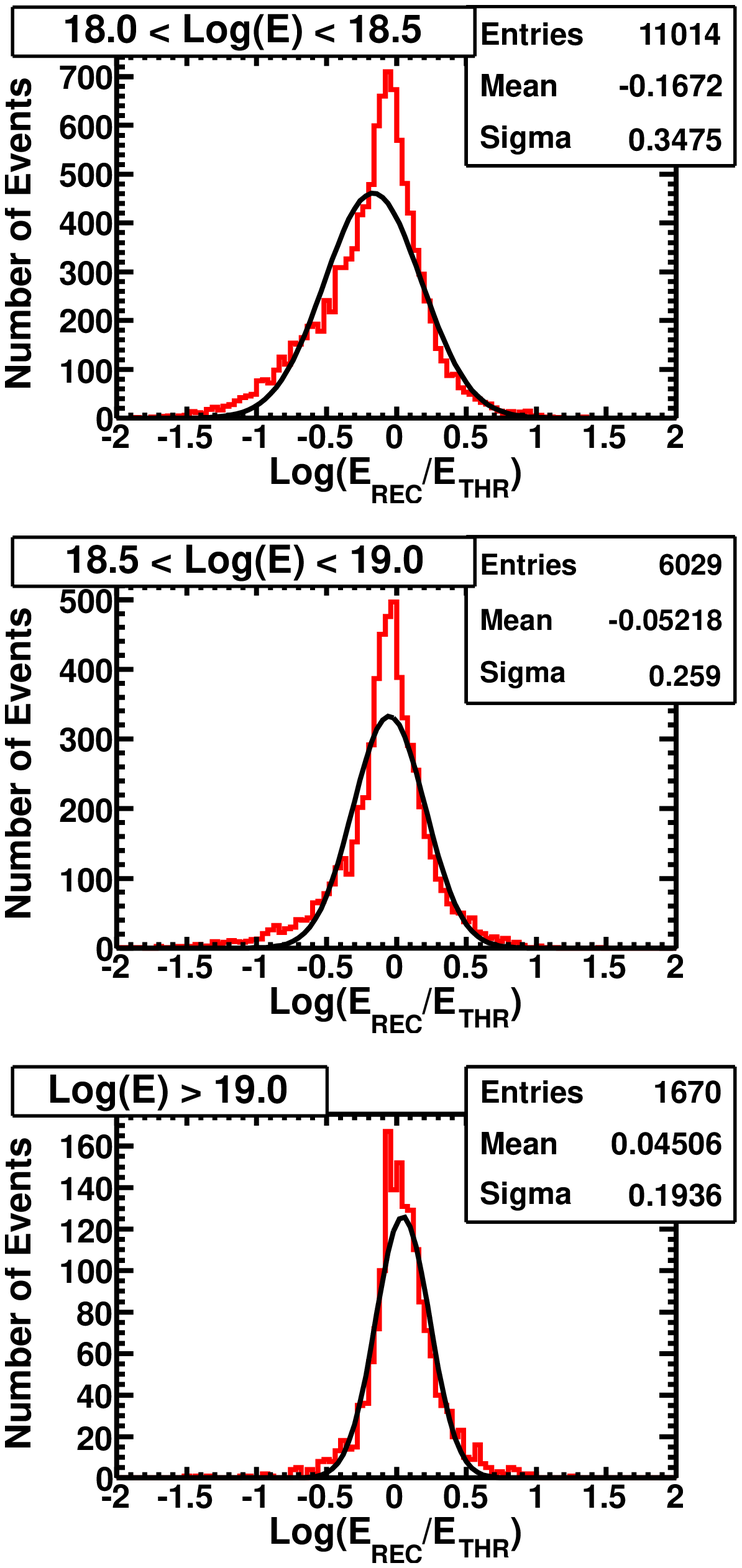}
}
\caption[Resolutions for Middle Drum monocular reconstructed energy: shown are events broken up into energy ranges: $10^{18}$ $<$ E $<$ $10^{18.5}$~eV (top),  $10^{18.5}$ $<$ E $<$ $10^{19.0}$~eV (middle), and E $>$ $10^{19.0}$~eV (bottom).]{Resolutions for Middle Drum monocular reconstructed energy: events are shown by energy range: $10^{18}$ $<$ E $<$ $10^{18.5}$~eV (top),  $10^{18.5}$ $<$ E $<$ $10^{19.0}$~eV (middle), and E $>$ $10^{19.0}$~eV (bottom). In each case, the red histogram shows the log of the ratio of the reconstructed and thrown energy for each event. The black line is a gaussian fit to the histogram. The energy resolutions (10\%, 7\%, and 6\%) for the hybrid reconstruction (in the previous figure) represent more than a factor of two improvement over the monocular reconstruction (34\%, 26\% and 19\%). Note that the horizontal scale is changed in the monocular case.}
\label{fig:FDresEng}
\end{figure}

Figure \ref{fig:resEng} shows the energy resolution for the MD hybrid reconstruction in three energy ranges. The improved geometrical resolution over the MD monocular measurement (Figure \ref{fig:resGeom}) directly contributes to the improvement in the energy resolution for the hybrid reconstruction. The resolution in energy starts at about 10\%  for the energy range of $10^{18.0}-10^{18.5}$ eV and improves with increased energy. This is more than a factor of two improvement over the MD monocular reconstruction, shown in Figure \ref{fig:FDresEng}. These improvements show the strength of the extra constraint of SD information. 

\subsection{Data/MC Comparisons}
MC simulations are also used to calculate the aperture of the TA detector which is then folded in with running time of each detector element to calculate the exposure. However, the MC must provide a faithful representation of distributions in the data for the aperture calculation and the resultant measured flux to be trusted. We validate the fidelity of the simulation by making a series of comparisons between the data and the Monte Carlo simulated data for a number of parameter distributions. In particular, we compare those variables directly connected to the aperture. 

Here we show the distributions from accepted events of both the data and MC, having been processed using the same analysis programs and subjected to the same selection cuts. In addition, for each comparison, a Kolmogorov-Smirnov (K-S) test is performed to compare the data and MC distributions. This test is appropriate for the small size of the data sample. In nearly every case, except when statistics are small (in the highest energy range), the agreement between data and MC for these parameters in these comparisons is very good. 

\begin{figure}[tbp]
  \centerline{
   \includegraphics[width=0.9\linewidth]{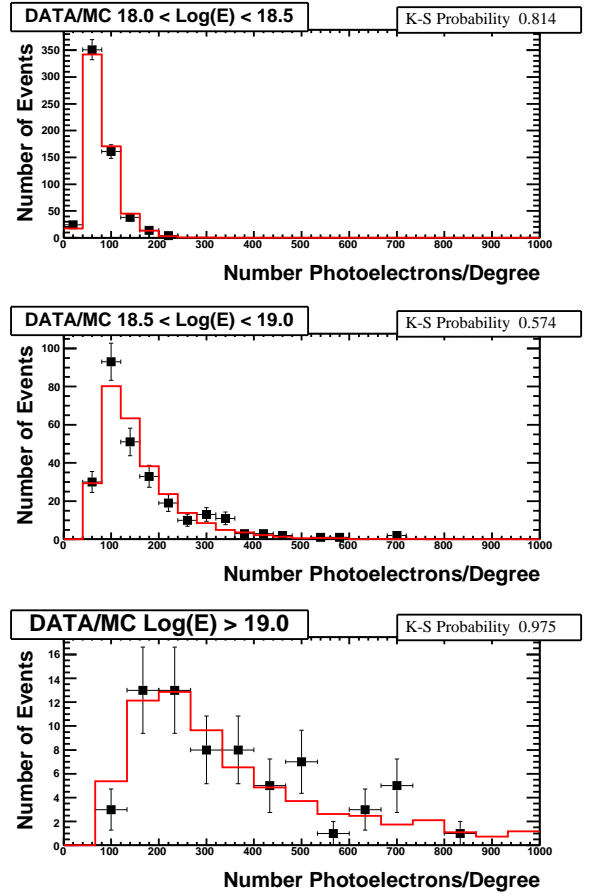}
   }
   \caption[Data-MC comparison: the Middle Drum hybrid number of photoelectrons per degree of track length is shown in three energy ranges: top to bottom,  $10^{18.0}$ $<$ E $<$ $10^{18.5}$~eV,  $10^{18.5}$ $<$ E $<$ $10^{19.0}$~eV, and E $>$ $10^{19.0}$~eV, respectively, to show the evolution of this parameter with energy.]{Data-MC comparison: the Middle Drum hybrid number of photoelectrons per degree of track length is shown in three energy ranges: top to bottom,  $10^{18.0}$ $<$ E $<$ $10^{18.5}$~eV,  $10^{18.5}$ $<$ E $<$ $10^{19.0}$~eV, and E $>$ $10^{19.0}$~eV, respectively, to show the evolution of this parameter with energy. The distribution of measurements is shown for the data (black points with error bars) and MC (red histogram). The MC has been normalized to the area of the data in these plots. This figure shows that the data and MC agreement for this parameter is not dependent on energy.} 
  \label{fig:dtmcNpeDeg3}
\end{figure}

Figure \ref{fig:dtmcNpeDeg3} shows the Data/MC comparisons for the number of photoelectrons per degree of track length of accepted events. The agreement here gives a good indication that the simulated detector response is accurate. The comparisons are shown in three energy regions.

\begin{figure}[tbp]
  \centerline{
   \includegraphics[width=0.9\linewidth]{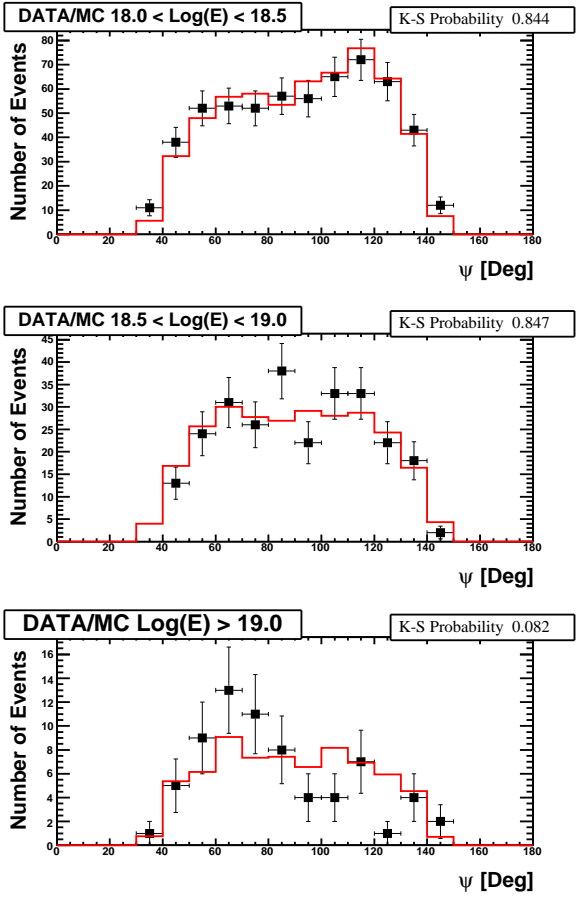}
   }
   \caption[Data-MC comparison: the Middle Drum hybrid in-plane angle ($\psi$) is shown in three energy ranges: top to bottom,  $10^{18.0}$ $<$ E $<$ $10^{18.5}$~eV,  $10^{18.5}$ $<$ E $<$ $10^{19.0}$~eV, and E $>$ $10^{19.0}$~eV, respectively, to show the evolution of this parameter with energy.]{Data-MC comparison: the Middle Drum hybrid in-plane angle ($\psi$) is shown in three energy ranges: top to bottom,  $10^{18.0}$ $<$ E $<$ $10^{18.5}$~eV,  $10^{18.5}$ $<$ E $<$ $10^{19.0}$~eV, and E $>$ $10^{19.0}$~eV, respectively, to show the evolution of this parameter with energy. The distribution of measurements is shown for the data (black points with error bars) and MC (red histogram). The MC has been normalized to the area of the data in these plots. This figure shows that the data and MC agreement for this parameter is not dependent on energy.} 
  \label{fig:dtmcPsi3}
\end{figure}

Figure \ref{fig:dtmcPsi3} shows the Data/MC comparisons for the in-plane angle ($\psi$) for showers in three energy ranges. This comparison shows whether we are simulating the evolution of this parameter reliably with energy. MD hybrid analysis is optimized in the region of $10^{18.5}$ ~eV to $10^{19.0}$~eV and therefore, the most accurate reconstructions of showers are found in this energy range. It is important to note, however, that the agreement between data and MC is well reproduced in all energy ranges.

\begin{figure}[tbp]
  \centerline{
   \includegraphics[width=0.9\linewidth]{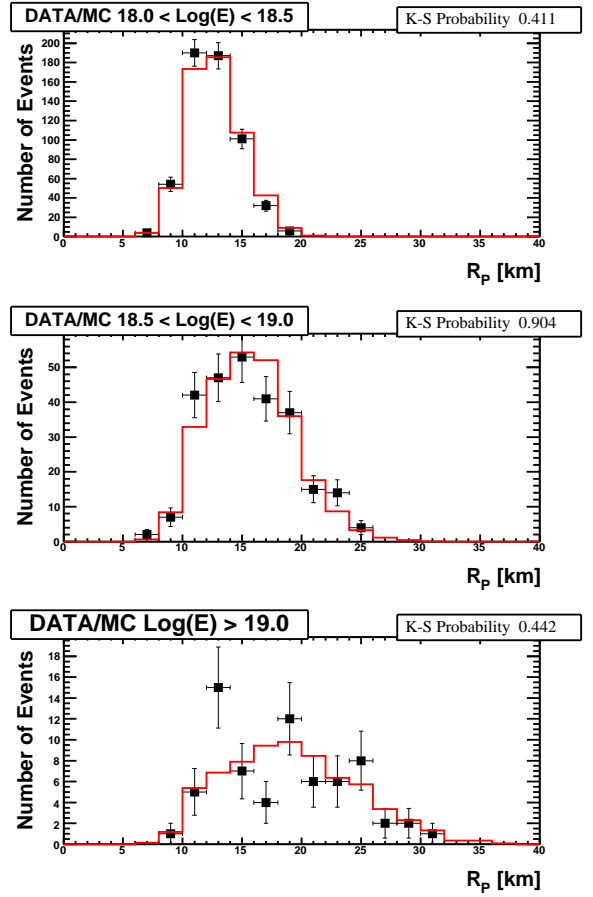}
   }
   \caption[Data-MC comparison: the Middle Drum hybrid impact parameter ($R_{P}$) is shown in three energy ranges: top to bottom,  $10^{18.0}$ $<$ E $<$ $10^{18.5}$~eV,  $10^{18.5}$ $<$ E $<$ $10^{19.0}$~eV, and E $>$ $10^{19.0}$~eV, respectively, to show the evolution of this parameter with energy.]{Data-MC comparison: the Middle Drum hybrid impact parameter ($R_{P}$) is shown in three energy ranges: top to bottom,  $10^{18.0}$ $<$ E $<$ $10^{18.5}$~eV,  $10^{18.5}$ $<$ E $<$ $10^{19.0}$~eV, and E $>$ $10^{19.0}$~eV, respectively, to show the evolution of this parameter with energy. The distribution of measurements is shown for the data (black points with error bars) and MC (red histogram). The MC has been normalized to the area of the data in these plots. This figure shows that the data and MC agreement for this parameter is not dependent on energy.} 
  \label{fig:dtmcRp3}
\end{figure}

\begin{figure}[tbp]
  \centerline{
   \includegraphics[width=0.9\linewidth]{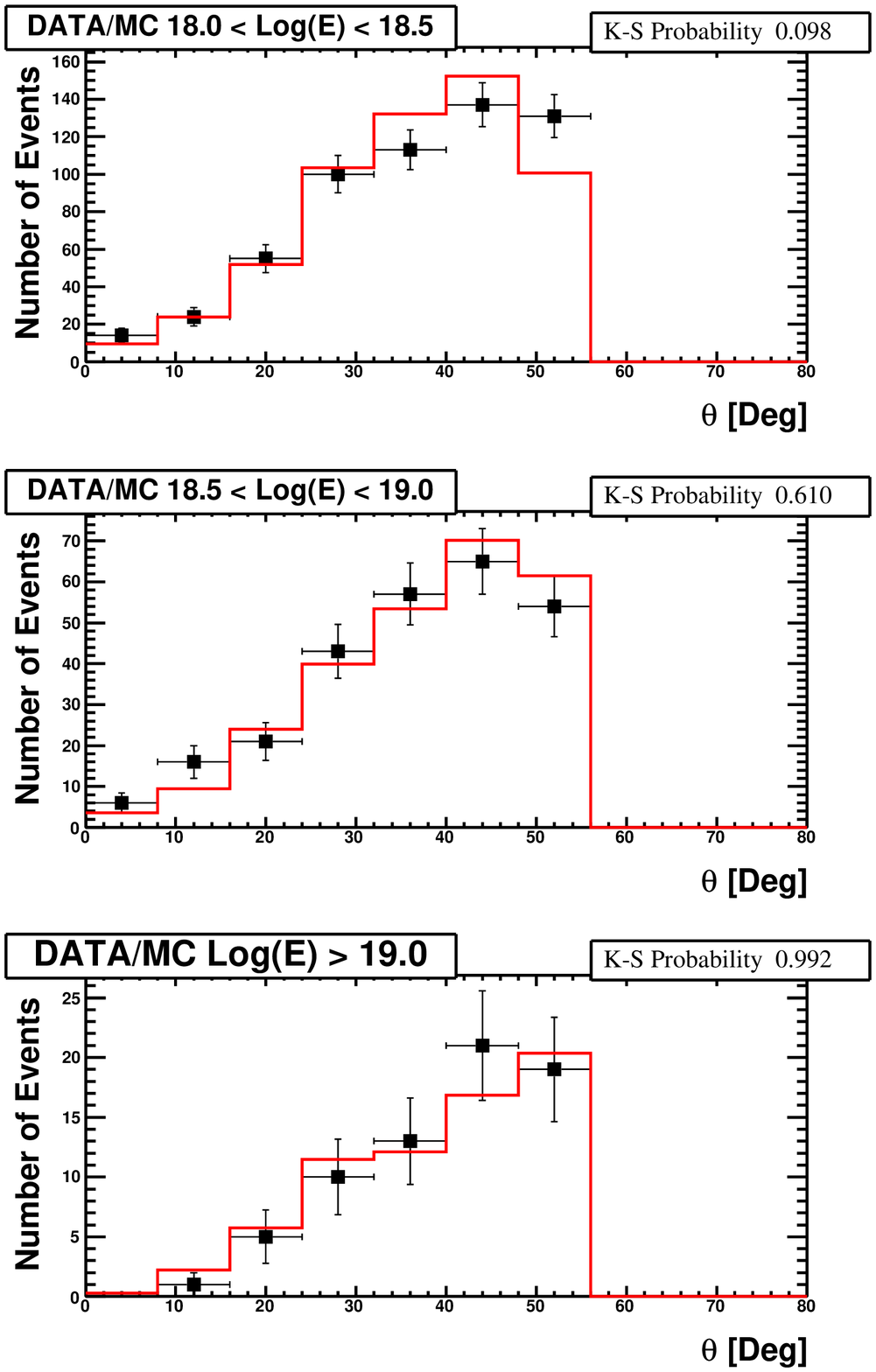}
   }
   \caption[Data-MC comparison: the Middle Drum hybrid zenith angle ($\theta$) is shown in three energy ranges: top to bottom,  $10^{18.0}$ $<$ E $<$ $10^{18.5}$~eV,  $10^{18.5}$ $<$ E $<$ $10^{19.0}$~eV, and E $>$ $10^{19.0}$~eV, respectively, to show the evolution of this parameter with energy.]{Data-MC comparison: the Middle Drum hybrid zenith angle ($\theta$) is shown in three energy ranges: top to bottom,  $10^{18.0}$ $<$ E $<$ $10^{18.5}$~eV,  $10^{18.5}$ $<$ E $<$ $10^{19.0}$~eV, and E $>$ $10^{19.0}$~eV, respectively, to show the evolution of this parameter with energy. The distribution of measurements is shown for the data (black points with error bars) and MC (red histogram). The MC has been normalized to the area of the data in these plots. This figure shows that the data and MC agreement for this parameter is not dependent on energy.} 
  \label{fig:dtmcZen3}
\end{figure}

\begin{figure}[tbp]
  \centerline{
   \includegraphics[width=0.9\linewidth]{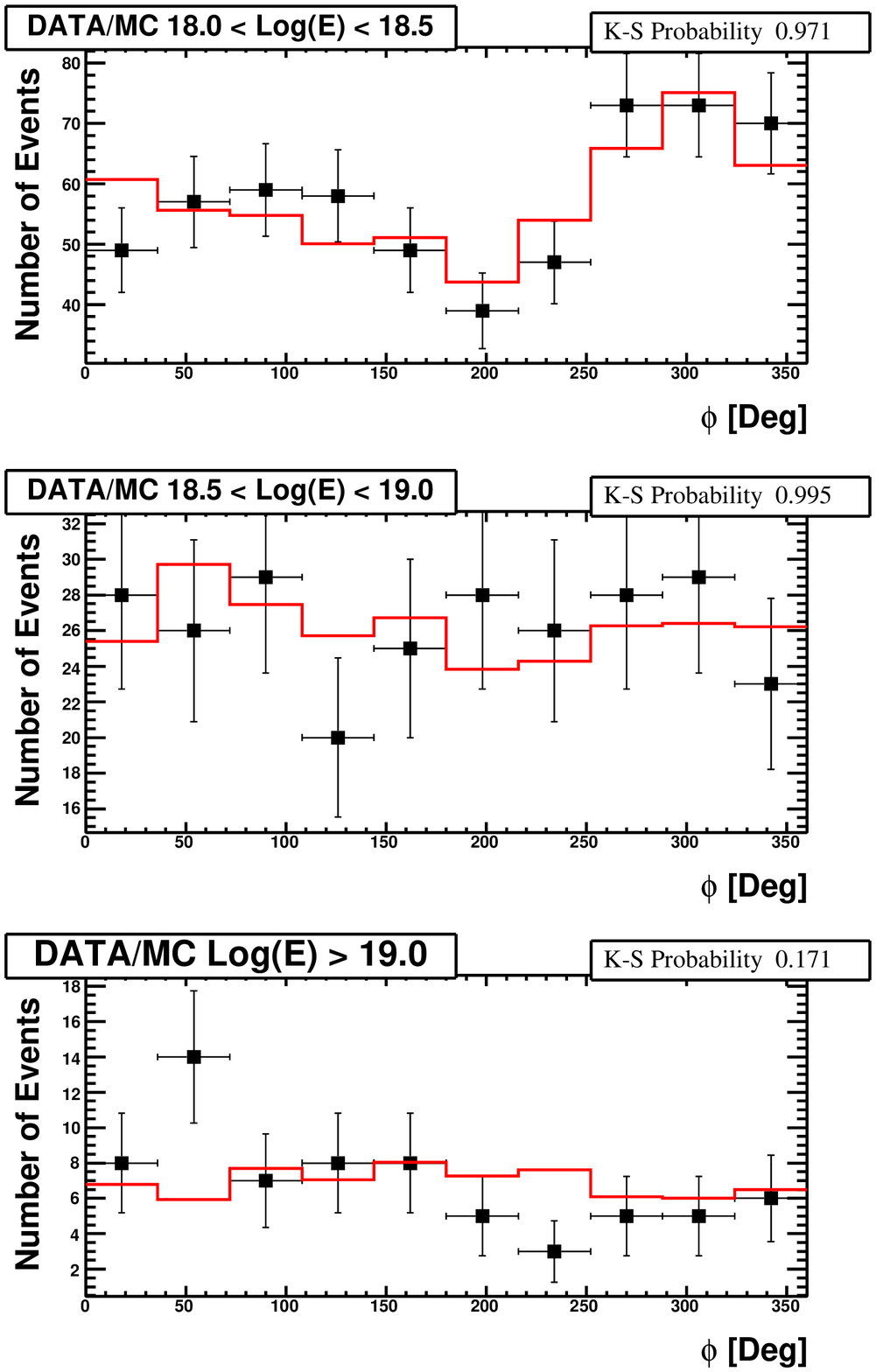}
   }
   \caption[Data-MC comparison: the Middle Drum hybrid azimuthal angle ($\phi$) (angle of the shower with respect to east is shown in three energy ranges: top to bottom,  $10^{18.0}$ $<$ E $<$ $10^{18.5}$~eV,  $10^{18.5}$ $<$ E $<$ $10^{19.0}$~eV, and E $>$ $10^{19.0}$~eV, respectively, to show the evolution of this parameter with energy.]{Data-MC comparison: the Middle Drum hybrid azimuthal angle ($\phi$) (angle of the shower with respect to east) is shown in three energy ranges: top to bottom,  $10^{18.0}$ $<$ E $<$ $10^{18.5}$~eV,  $10^{18.5}$ $<$ E $<$ $10^{19.0}$~eV, and E $>$ $10^{19.0}$~eV, respectively, to show the evolution of this parameter with energy. The distribution of measurements is shown for the data (black points with error bars) and MC (red histogram). The MC has been normalized to the area of the data in these plots. This figure shows that the data and MC agreement for this parameter is not dependent on energy.} 
  \label{fig:dtmcPhi3}
\end{figure}

Figures \ref{fig:dtmcRp3}, \ref{fig:dtmcZen3}, and \ref{fig:dtmcPhi3} show Data/MC comparisons for the impact parameter ($R_{P}$), zenith angle ($\theta$) and azimuthal angle ($\phi$) for three energy ranges. Again, the agreement between data and MC is consistently excellent in all three ranges in these plots. The K-S probability for each comparison is shown on the plot and indicates good agreement. 

\section{Middle Drum Hybrid Energy Spectrum}
The energy spectrum refers to the differential flux of cosmic rays. It is calculated by taking the number of data events per energy bin and dividing by the exposure and energy interval for that bin, as shown in Equation \ref{eq:Spectrum}.
\begin{equation}
J(E) = \frac{N(E)}{ A\Omega(E) \times \Delta t \times \Delta E}
\label{eq:Spectrum}
\end{equation}

\noindent Here, $N(E)$ refers to the number of reconstructed events in an energy bin, $A\Omega$ is the calculated aperture for the energy bin, $\Delta t$ is the hybrid detector on-time, and $\Delta E$ is the energy interval covered by the bin. The systematic uncertainty of the energy calculation due to atmospheric conditions was taken into account when calculating this flux. A study of the vertical aerosol optical depth found that uncertainty is $\sim3$\% \cite{Rodriguez2011}.

The exposure is calculated by taking the aperture per energy bin and multiplying by the on-time for the detector. The SD array collects data 24~hours a day. Taking into account the data acquisition system and the individual detectors in the array that are not working periodically, the array has better than 95\% on-time. Therefore, the main contribution to the on-time calculation for this analysis comes from the fluorescence detector. The MD detector only operates on clear, moonless nights, with a minimum of three hours of dark time. 

\begin{figure}[tbp]
  \centerline
  {
   \includegraphics[width=\linewidth]{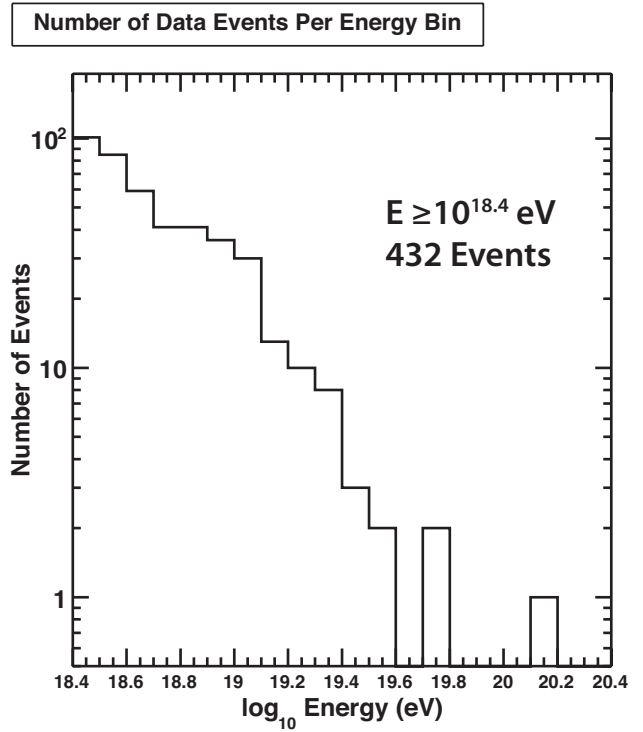}
   }
 \caption{The raw energy distribution of events passing all quality cuts observed in hybrid mode by the Middle Drum telescope site: the events are binned in energy. a total of 432 events remain that were used to calculate the MD hybrid spectrum.}
  \label{fig:NumEvents}
\end{figure}

The MD hybrid energy spectrum was calculated using four years of data, the number of integrated good weather on-time hours was 3071.8 between May 11, 2008 (SD turn on) and May 19, 2012. Good weather data includes only data taken on nights when clouds were not present in the directions that the MD telescopes point, namely, South and East. There were 1580 triggered events in the data set. After taking dark time and weather cuts into account, the MD detector duty cycle is $\sim$9\%. The final data set has 432 events with reconstructed energies above $10^{18.4}$~eV, below which, the hybrid detector aperture drops off steeply. The raw energy distribution of these events is shown in Figure \ref{fig:NumEvents}. Note that the highest energy event has a reconstructed energy of 1.32$\times$10$^{20}$~eV. This event was not used in the SD monocular spectrum because it was reconstructed with a zenith angle of 55.7$\Deg$, and events with zenith angle $>$45$\Deg$ were cut from that analysis due to uncertainty in reconstructing the event energy using only the SD array.

\begin{figure}[tbp]
  \centerline
  {
   \includegraphics[width=\linewidth]{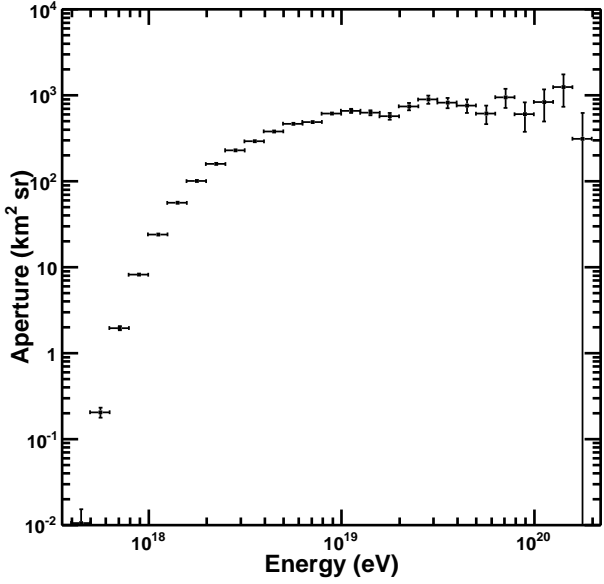}
   }
 \caption{The calculated Middle Drum hybrid aperture from proton Monte Carlo.}
  \label{fig:Aperture}
\end{figure}

Figure \ref{fig:Aperture} shows the calculated aperture from the Monte Carlo. The aperture falls off steeply below $10^{18.4}$ eV. Therefore, for the purpose of this analysis, the spectrum is calculated and shown for energies of $10^{18.4}$ eV and above. 

\begin{figure}[tbp]
  \centerline
  {
   \includegraphics[width=1.0\linewidth]{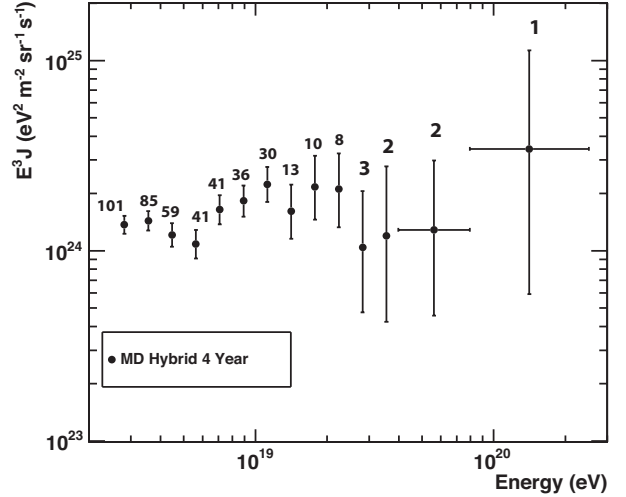}
   }
 \caption[Middle Drum hybrid energy spectrum: shown is the differential flux of ultra high energy cosmic rays as a function of energy.]{Middle Drum hybrid 4 year energy spectrum: shown is the differential flux of ultra high energy cosmic rays with energies, $10^{18.4}<E<10^{20.2}$~eV, as a function of energy. The flux has been multiplied by a factor of $E^{3}$ to take out the steep slope of the overall spectrum and better show the fine structure. The numbers above the data points indicate the number of observed events in those bins. Note that the top energy bins have been combined due to small statistics. Only three events were observed in hybrid mode with energies $>10^{19.6}$~eV.}
  \label{fig:HBSpecOnly}
\end{figure}

Figure \ref{fig:HBSpecOnly} shows the differential flux as a function of energy for the MD hybrid events. Due to the geometric and temporal limitations of collecting data in hybrid mode, the statistics for this spectrum are relatively small. 

The MD hybrid analysis plays an important role in connecting the measurements of the High Resolution Fly's Eye (HiRes) experiment to the Telescope Array experiment. The MD monocular spectrum \cite{Rodriguez2012} provided the retrograde link between the TA and HiRes spectra, and this hybrid analysis takes this link a step further by creating a direct connection between the MD detector and the SD array. For this purpose, comparisons of the measured energy and energy spectrum with other TA analyses are discussed in the next section.

\section{Comparison to MD Monocular and SD Spectra}
\begin{figure}[tbp]
  \centerline
  {
   \includegraphics[width=0.9\linewidth]{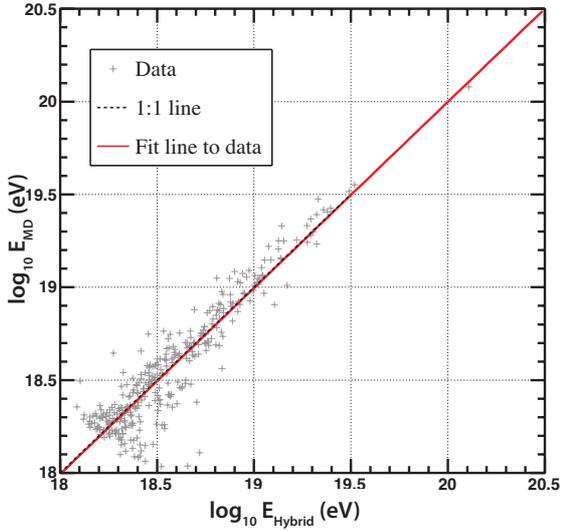}
   }
 \caption[A scatter plot showing the event-by-event energy comparison of data events reconstructed by the Middle Drum hybrid analysis (X-axis), and the Middle Drum monocular analysis (Y-axis).]{A scatter plot showing the event-by-event comparison of the energies of data events reconstructed by the Middle Drum hybrid analysis (X-axis), and those by the Middle Drum monocular analysis (Y-axis). The dashed line indicates the 1:1 line, while the solid line is a fit to the data.}
  \label{fig:HybridMDHScat}
\end{figure}

\begin{figure}[tbp]
  \centerline
  {
   \includegraphics[width=0.9\linewidth]{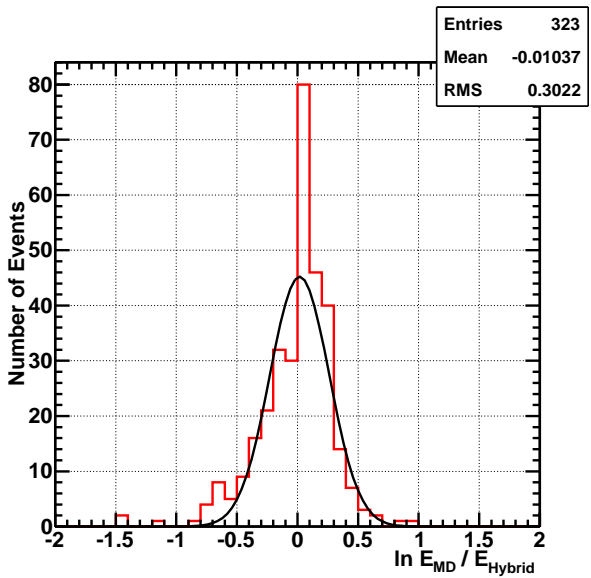}
   }
 \caption{A histogram of the log ratio of the energies of events reconstructed by the Middle Drum monocular analysis over those by the Middle Drum hybrid analysis: the width in this histogram is dominated by the resolution in the MD monocular reconstruction ($\sim$26\%).}
  \label{fig:HybridMDHHist}
\end{figure}

An event-by-event study was performed comparing the MD monocular data to the MD hybrid data. Figure \ref{fig:HybridMDHScat} shows the energy reconstruction comparison. The systematic uncertainties in the MD monocular spectrum are primarily due to atmospheric changes, which are the same for the hybrid detector. The dashed line in the figure is the 1:1 line, while the solid line represents a fit to the data. No statistically significant bias is seen here. Furthermore, Figure \ref{fig:HybridMDHHist} shows a histogram of the log ratio of the MD monocular reconstructed energy over the MD hybrid reconstructed energy. Again, no bias is seen. 

\begin{figure}[tbp]
  \centerline
  {
   \includegraphics[width=1.0\linewidth]{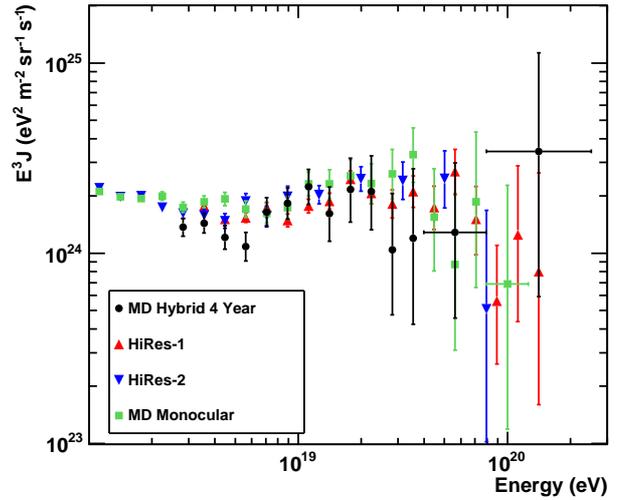}
   }
 \caption[The Middle Drum hybrid energy spectrum (black circles) compared with the spectrum measured by the Middle Drum detector in monocular mode (green squares), as well as the spectra measured by the HiRes-1 (red triangles) and HiRes-2 (blue triangles) detectors.]{The Middle Drum hybrid energy spectrum (black circles) compared with the spectrum measured by the Middle Drum detector in monocular mode (green squares), as well as the spectra measured by the HiRes-1 (red triangles) and HiRes-2 (blue triangles) detectors.}
  \label{fig:SpectraHBHiResH}
\end{figure}

Figure \ref{fig:SpectraHBHiResH} compares the MD monocular spectrum with this MD hybrid analysis, as well as the HiRes-1 and -2 spectra. The MD hybrid spectrum is in reasonable agreement with the MD monocular spectrum as well as both of the HiRes spectra (see table \ref{tab:chi2test}). 

\begin{figure}[tbp]
  \centerline
  {
   \includegraphics[width=0.9\linewidth]{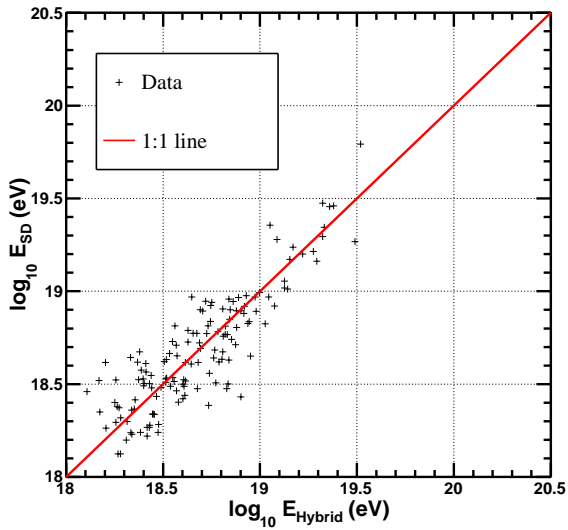}
   }
 \caption[A scatter plot showing the event-by-event comparison of the reconstructed event energy by the Middle Drum hybrid analysis (X-axis), and the Surface Detector analysis (Y-axis).]{A scatter plot showing the event-by-event comparison of the reconstructed event energy by the Middle Drum hybrid analysis (X-axis), and the Surface Detector analysis (Y-axis). The line indicates the 1:1 line.}
  \label{fig:HybridSDScat}
\end{figure}

\begin{figure}[tbp]
  \centerline
  {
   \includegraphics[width=0.9\linewidth]{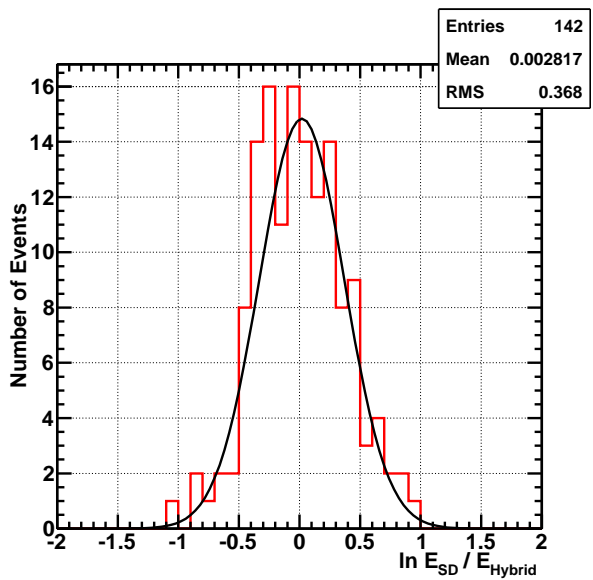}
   }
 \caption{A histogram of the log ratio of the energies of events reconstructed by the Surface Detector analysis over those by the Middle Drum hybrid analysis: the width in this histogram is dominated by the resolution in the SD reconstruction ($\sim$29\%).}
  \label{fig:HybridSDHist}
\end{figure}

\begin{figure}[tbp]
  \centerline
  {
   \includegraphics[width=1.0\linewidth]{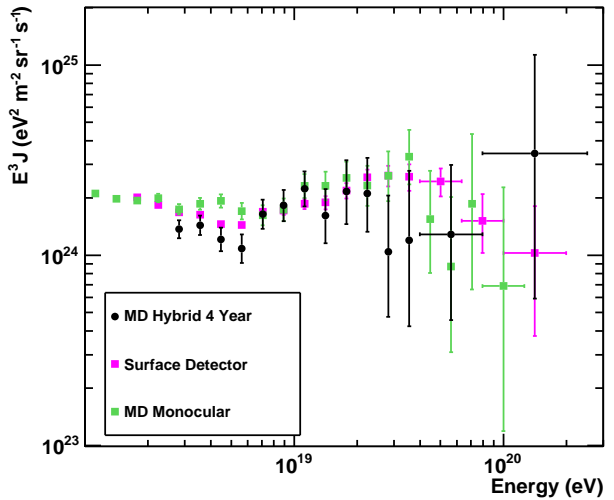}
   }
 \caption[The Middle Drum hybrid spectrum (black circles) compared with the spectrum measured by the surface array (purple squares): the MD monocular spectrum (green squares) is shown for reference.]{The Middle Drum hybrid spectrum (black circles) compared with the spectrum measured by the surface array (purple squares): the MD monocular spectrum (green squares) is shown for reference.}
  \label{fig:SpectraHBSD}
\end{figure}

The next step in linking the HiRes spectrum to the Telescope Array is a comparison between the MD hybrid energy spectrum and that measured by the TA SD. Event-by-event comparisons were also made between the hybrid and the SD measurement. SD event energies are estimated using the correlation of the number of particles at a point 800~m from the shower core, S800, and the zenith angle of the event with the primary energy from the MC study. A comparison of TA FD and SD events found that the CORSIKA simulated showers were producing higher than expected numbers of particles at S800. Therefore, a scaling factor of 1.27 was used to calculate the SD energies \cite{Ivanov2012,Jui2012}. Figure \ref{fig:HybridSDScat} shows the scatter plot of the MD hybrid reconstructed energy of each event vs the SD reconstructed energy. Again, the 1:1 line is shown, and there is no significant bias in the data. The histogram of the log ratio of the SD monocular reconstructed energy over the MD hybrid reconstructed energy is shown in figure \ref{fig:HybridSDHist}. And finally, the MD hybrid spectrum is shown in comparison to the SD spectrum in Figure \ref{fig:SpectraHBSD}. They are in good agreement (see table \ref{tab:chi2test}).

For each comparison, a $\chi^2$ test was performed to see how well the spectra agree. The results of the comparison of this hybrid analysis with each of the other spectra are summarized in Table \ref{tab:chi2test}. 


\begin{table}[h]
\centering
\caption[A summary of the results of a $\chi^2$ test performed to compare four analyses with the Middle Drum hybrid analysis is given.]{A summary of the results of a $\chi^2$ test performed to compare four analyses with the Middle Drum hybrid analysis is given. The comparisons were performed for each analysis using only bins with energy $18.4<log_{10}(E) < 19.4$.} 
\vspace{10pt} 
\label{tab:chi2test}
\begin{tabular}{l  c  c  c  c }
\firsthline
\noalign{\smallskip}
Data & Energy Range & $\chi^2$ & $\#$ Degrees \\ 
&log$_{10} (E)$ & &of Freedom \\
\noalign{\smallskip}
\hline \hline
\noalign{\smallskip}
MD Mono & 18.4-19.4 & 23.78 & 10 \\
\noalign{\smallskip}
\hline
\noalign{\smallskip}
SD Mono & 18.4-19.4 & 10.56 & 10 \\
\noalign{\smallskip}
\hline
\noalign{\smallskip}
HiRes-1 & 18.5-19.4 & 16.65 & 9 \\
\noalign{\smallskip}
\hline
\noalign{\smallskip}
HiRes-2 & 18.4-19.0 & 19.30 & 6 \\
\noalign{\smallskip}
\hline
\end{tabular} 
\end{table}

\section{Conclusion}
In conclusion, we measure the hybrid energy spectrum using the MD detector in conjunction with the SD. The MD site re-utilizes the telescopes and electronics from the HiRes experiment. Therefore, this work directly links the measurements of these two experiments. The MD monocular spectrum has been shown previously to agree with the HiRes spectra. This hybrid analysis establishes a starting point for comparison between HiRes and TA spectra. The MD hybrid spectrum is in good agreement with the MD monocular spectrum and the HiRes spectra, confirming this result. Furthermore, the hybrid spectrum agrees with the SD monocular spectrum, confirming the HiRes result from the perspective of the TA experiment as a whole.

\include{TAacknowledgments-20140620}
\newpage
\bibliographystyle{model1-num-names}
\bibliography{library}

\end{document}

%% file: TAauthor_list-20140620_formatted.tex
\author[1]{R.U.~Abbasi}
\author[13]{M.~Abe}
\author[1]{T.Abu-Zayyad}
\author[1]{M.G.~Allen\corref{cor1}}
\ead{monica@cosmic.utah.edu}
\author[1]{R.~Anderson}
\author[2]{R.~Azuma}
\author[1]{E.~Barcikowski}
\author[1]{J.W.~Belz}
\author[1]{D.R.~Bergman}
\author[1]{S.A.~Blake}
% ...

\address[1]{High Energy Astrophysics Institute and Department of Physics and Astronomy, University of Utah, Salt Lake City, Utah, USA}
\address[2]{Graduate School of Science and Engineering, Tokyo Institute of Technology, Meguro, Tokyo, Japan}
% ...
\address[13]{The Graduate School of Science and Engineering, Saitama University, Saitama, Saitama, Japan}
% ...
\cortext[cor1]{Corresponding Author}

%% file: abstract.tex
\begin{abstract} The Telescope Array experiment studies ultra high energy cosmic rays using a hybrid detector. Fluorescence telescopes measure the longitudinal development of the extensive air shower generated when a primary cosmic ray particle interacts with the atmosphere. Meanwhile, scintillator detectors measure the lateral distribution of secondary shower particles that hit the ground. The Middle Drum (MD) fluorescence telescope station consists of 14 telescopes from the High Resolution Fly's Eye (HiRes) experiment, providing a direct link back to the HiRes measurements. Using the scintillator detector data in conjunction with the telescope data improves the geometrical reconstruction of the showers significantly, and hence, provides a more accurate reconstruction of the energy of the primary particle. The Middle Drum hybrid spectrum is presented and compared to that measured by the Middle Drum station in monocular mode. Further, the hybrid data establishes a link between the Middle Drum data and the surface array. A comparison between the Middle Drum hybrid energy spectrum and scintillator Surface Detector (SD) spectrum is also shown.
\end{abstract}

%% file: TAacknowledgments-20140620.tex
\section*{Acknowledgment}
The Telescope Array experiment is supported by the Japan Society for
the Promotion of Science through Grants-in-Aids for Scientific
Research on Specially Promoted Research (21000002) ``Extreme Phenomena
in the Universe Explored by Highest Energy Cosmic Rays'' and for
Scientific Research (19104006), and the Inter-University Research
Program of the Institute for Cosmic Ray Research; by the U.S. National
Science Foundation awards PHY-0307098, PHY-0601915, PHY-0649681,
PHY-0703893, PHY-0758342, PHY-0848320, PHY-1069280, and PHY-1069286;
by the National Research Foundation of Korea (2007-0093860, R32-10130,
2012R1A1A2008381, 2013004883); by the Russian Academy of Sciences,
RFBR grants 11-02-01528a and 13-02-01311a (INR), IISN project
No. 4.4509.10 and Belgian Science Policy under IUAP VII/37 (ULB). The
foundations of Dr. Ezekiel R. and Edna Wattis Dumke, Willard L. Eccles
and the George S. and Dolores Dore Eccles all helped with generous
donations. The State of Utah supported the project through its
Economic Development Board, and the University of Utah through the
Office of the Vice President for Research. The experimental site
became available through the cooperation of the Utah School and
Institutional Trust Lands Administration (SITLA), U.S. Bureau of Land
Management, and the U.S. Air Force. We also wish to thank the people
and the officials of Millard County, Utah for their steadfast and warm
support. We gratefully acknowledge the contributions from the
technical staffs of our home institutions. An allocation of computer
time from the Center for High Performance Computing at the University
of Utah is gratefully acknowledged.

%% file: MDHybrid4YSpectrumELS_arXiv.bbl
\begin{thebibliography}{17}
\expandafter\ifx\csname natexlab\endcsname\relax\def\natexlab#1{#1}\fi
\providecommand{\bibinfo}[2]{#2}
\ifx\xfnm\relax \def\xfnm[#1]{\unskip,\space#1}\fi
\bibitem[{Abu-Zayyad et~al.(2012{\natexlab{a}})Abu-Zayyad, Aida, Allen,
  Anderson, Azuma, Barcikowski, Belz, Bergman, Blake, Cady, Cheon, Chiba,
  Chikawa, Cho, Cho, Fujii, Fujii, Fukuda, Fukushima, Gorbunov, Hanlon,
  Hayashi, Hayashi, Hayashida, Hibino, Hiyama, Honda, Iguchi, Ikeda, Ikuta,
  Inoue, Ishii, Ishimori, Ivanov, Iwamoto, Jui, Kadota, Kakimoto, Kalashev,
  Kanbe, Kasahara, Kawai, Kawakami, Kawana, Kido, Kim, Kim, Kim, Kitamoto,
  Kitamura, Kitamura, Kobayashi, Kobayashi, Kondo, Kuramoto, Kuzmin, Kwon, Lim,
  Machida, Martens, Martineau, Matsuda, Matsuura, Matsuyama, Matthews,
  Minamino, Miyata, Murano, Nagataki, Nakamura, Nam, Nonaka, Ogio, Ohnishi,
  Ohoka, Oki, Oku, Okuda, Oshima, Ozawa, Park, Pshirkov, Rodriguez, Roh,
  Rubtsov, Ryu, Sagawa, Sakurai, Sampson, Scott, Shah, Shibata, Shibata,
  Shimodaira, Shin, Shin, Shirahama, Smith, Sokolsky, Sonley, Springer, Stokes,
  Stratton, Stroman, Suzuki, Takahashi, Takeda, Taketa, Takita, Tameda, Tanaka,
  Tanaka, Tanaka, Thomas, Thomson, Tinyakov, Tkachev, Tokuno, Tomida, Troitsky,
  Tsunesada, Tsutsumi, Tsuyuguchi, Uchihori, Udo, Ukai, Vasiloff, Wada, Wong,
  Wood, Yamakawa, Yamane, Yamaoka, Yamazaki, Yang, Yoneda, Yoshida, Yoshii,
  Zollinger, and Zundel}]{Rodriguez2012}
\bibinfo{author}{T.~Abu-Zayyad}, \bibinfo{author}{R.~Aida},
  \bibinfo{author}{M.~Allen}, \bibinfo{author}{R.~Anderson},
  \bibinfo{author}{R.~Azuma}, \bibinfo{author}{E.~Barcikowski},
  \bibinfo{author}{J.~W. Belz}, \bibinfo{author}{D.~R. Bergman},
  \bibinfo{author}{S.~A. Blake}, \bibinfo{author}{R.~Cady},
  \bibinfo{author}{B.~G. Cheon}, \bibinfo{author}{J.~Chiba},
  \bibinfo{author}{M.~Chikawa}, \bibinfo{author}{E.~J. Cho},
  \bibinfo{author}{W.~R. Cho}, \bibinfo{author}{H.~Fujii},
  \bibinfo{author}{T.~Fujii}, \bibinfo{author}{T.~Fukuda},
  \bibinfo{author}{M.~Fukushima}, \bibinfo{author}{D.~Gorbunov},
  \bibinfo{author}{W.~Hanlon}, \bibinfo{author}{K.~Hayashi},
  \bibinfo{author}{Y.~Hayashi}, \bibinfo{author}{N.~Hayashida},
  \bibinfo{author}{K.~Hibino}, \bibinfo{author}{K.~Hiyama},
  \bibinfo{author}{K.~Honda}, \bibinfo{author}{T.~Iguchi},
  \bibinfo{author}{D.~Ikeda}, \bibinfo{author}{K.~Ikuta},
  \bibinfo{author}{N.~Inoue}, \bibinfo{author}{T.~Ishii},
  \bibinfo{author}{R.~Ishimori}, \bibinfo{author}{D.~Ivanov},
  \bibinfo{author}{S.~Iwamoto}, \bibinfo{author}{C.~C.~H. Jui},
  \bibinfo{author}{K.~Kadota}, \bibinfo{author}{F.~Kakimoto},
  \bibinfo{author}{O.~Kalashev}, \bibinfo{author}{T.~Kanbe},
  \bibinfo{author}{K.~Kasahara}, \bibinfo{author}{H.~Kawai},
  \bibinfo{author}{S.~Kawakami}, \bibinfo{author}{S.~Kawana},
  \bibinfo{author}{E.~Kido}, \bibinfo{author}{H.~B. Kim},
  \bibinfo{author}{H.~K. Kim}, \bibinfo{author}{J.~H. Kim},
  \bibinfo{author}{K.~Kitamoto}, \bibinfo{author}{S.~Kitamura},
  \bibinfo{author}{Y.~Kitamura}, \bibinfo{author}{K.~Kobayashi},
  \bibinfo{author}{Y.~Kobayashi}, \bibinfo{author}{Y.~Kondo},
  \bibinfo{author}{K.~Kuramoto}, \bibinfo{author}{V.~Kuzmin},
  \bibinfo{author}{Y.~J. Kwon}, \bibinfo{author}{S.~I. Lim},
  \bibinfo{author}{S.~Machida}, \bibinfo{author}{K.~Martens},
  \bibinfo{author}{J.~Martineau}, \bibinfo{author}{T.~Matsuda},
  \bibinfo{author}{T.~Matsuura}, \bibinfo{author}{T.~Matsuyama},
  \bibinfo{author}{J.~N. Matthews}, \bibinfo{author}{M.~Minamino},
  \bibinfo{author}{K.~Miyata}, \bibinfo{author}{Y.~Murano},
  \bibinfo{author}{S.~Nagataki}, \bibinfo{author}{T.~Nakamura},
  \bibinfo{author}{S.~W. Nam}, \bibinfo{author}{T.~Nonaka},
  \bibinfo{author}{S.~Ogio}, \bibinfo{author}{M.~Ohnishi},
  \bibinfo{author}{H.~Ohoka}, \bibinfo{author}{K.~Oki},
  \bibinfo{author}{D.~Oku}, \bibinfo{author}{T.~Okuda},
  \bibinfo{author}{A.~Oshima}, \bibinfo{author}{S.~Ozawa},
  \bibinfo{author}{I.~H. Park}, \bibinfo{author}{M.~S. Pshirkov},
  \bibinfo{author}{D.~C. Rodriguez}, \bibinfo{author}{S.~Y. Roh},
  \bibinfo{author}{G.~Rubtsov}, \bibinfo{author}{D.~Ryu},
  \bibinfo{author}{H.~Sagawa}, \bibinfo{author}{N.~Sakurai},
  \bibinfo{author}{A.~L. Sampson}, \bibinfo{author}{L.~M. Scott},
  \bibinfo{author}{P.~D. Shah}, \bibinfo{author}{F.~Shibata},
  \bibinfo{author}{T.~Shibata}, \bibinfo{author}{H.~Shimodaira},
  \bibinfo{author}{B.~K. Shin}, \bibinfo{author}{J.~I. Shin},
  \bibinfo{author}{T.~Shirahama}, \bibinfo{author}{J.~D. Smith},
  \bibinfo{author}{P.~Sokolsky}, \bibinfo{author}{T.~J. Sonley},
  \bibinfo{author}{R.~W. Springer}, \bibinfo{author}{B.~T. Stokes},
  \bibinfo{author}{S.~R. Stratton}, \bibinfo{author}{T.~Stroman},
  \bibinfo{author}{S.~Suzuki}, \bibinfo{author}{Y.~Takahashi},
  \bibinfo{author}{M.~Takeda}, \bibinfo{author}{A.~Taketa},
  \bibinfo{author}{M.~Takita}, \bibinfo{author}{Y.~Tameda},
  \bibinfo{author}{H.~Tanaka}, \bibinfo{author}{K.~Tanaka},
  \bibinfo{author}{M.~Tanaka}, \bibinfo{author}{S.~B. Thomas},
  \bibinfo{author}{G.~B. Thomson}, \bibinfo{author}{P.~Tinyakov},
  \bibinfo{author}{I.~Tkachev}, \bibinfo{author}{H.~Tokuno},
  \bibinfo{author}{T.~Tomida}, \bibinfo{author}{S.~Troitsky},
  \bibinfo{author}{Y.~Tsunesada}, \bibinfo{author}{K.~Tsutsumi},
  \bibinfo{author}{Y.~Tsuyuguchi}, \bibinfo{author}{Y.~Uchihori},
  \bibinfo{author}{S.~Udo}, \bibinfo{author}{H.~Ukai},
  \bibinfo{author}{G.~Vasiloff}, \bibinfo{author}{Y.~Wada},
  \bibinfo{author}{T.~Wong}, \bibinfo{author}{M.~Wood},
  \bibinfo{author}{Y.~Yamakawa}, \bibinfo{author}{R.~Yamane},
  \bibinfo{author}{H.~Yamaoka}, \bibinfo{author}{K.~Yamazaki},
  \bibinfo{author}{J.~Yang}, \bibinfo{author}{Y.~Yoneda},
  \bibinfo{author}{S.~Yoshida}, \bibinfo{author}{H.~Yoshii},
  \bibinfo{author}{R.~Zollinger}, \bibinfo{author}{Z.~Zundel},
\newblock \bibinfo{title}{{The Energy Spectrum of Telescope Array's Middle Drum
  Detector and the Direct Comparison to the High Resolution Fly's Eye
  Experiment}},
\newblock \bibinfo{journal}{Astroparticle Physics} \bibinfo{volume}{39-40}
  (\bibinfo{year}{2012}{\natexlab{a}}) \bibinfo{pages}{109--119}.
\bibitem[{Abu-Zayyad et~al.(2012{\natexlab{b}})Abu-Zayyad, Aida, Allen,
  Anderson, Azuma, Barcikowski, Belz, Bergman, Blake, Cady, Cheon, Chiba,
  Chikawa, Cho, Cho, Fujii, Fujii, Fukuda, Fukushima, Gorbunov, Hanlon,
  Hayashi, Hayashi, Hayashida, Hibino, Hiyama, Honda, Iguchi, Ikeda, Ikuta,
  Inoue, Ishii, Ishimori, Ivanov, Iwamoto, Jui, Kadota, Kakimoto, Kalashev,
  Kanbe, Kasahara, Kawai, Kawakami, Kawana, Kido, Kim, Kim, Kim, Kitamoto,
  Kobayashi, Kobayashi, Kondo, Kuramoto, Kuzmin, Kwon, Lim, Machida, Martens,
  Martineau, Matsuda, Matsuura, Matsuyama, Matthews, Myers, Minamino, Miyata,
  Miyauchi, Murano, Nakamura, Nam, Nonaka, Ogio, Ohnishi, Ohoka, Oki, Oku,
  Okuda, Oshima, Ozawa, Park, Pshirkov, Rodriguez, Roh, Rubtsov, Ryu, Sagawa,
  Sakurai, Sampson, Scott, Shah, Shibata, Shibata, Shimodaira, Shin, Shin,
  Shirahama, Smith, Sokolsky, Sonley, Springer, Stokes, Stratton, Stroman,
  Suzuki, Takahashi, Takeda, Taketa, Takita, Tameda, Tanaka, Tanaka, Tanaka,
  Thomas, Thomson, Tinyakov, Tkachev, Tokuno, Tomida, Troitsky, Tsunesada,
  Tsutsumi, Tsuyuguchi, Uchihori, Udo, Ukai, Vasiloff, Wada, Wong, Wood,
  Yamakawa, Yamaoka, Yamazaki, Yang, Yoshida, Yoshii, Zollinger, and
  Zundel}]{Abu-Zayyad2012a}
\bibinfo{author}{T.~Abu-Zayyad}, \bibinfo{author}{R.~Aida},
  \bibinfo{author}{M.~Allen}, \bibinfo{author}{R.~Anderson},
  \bibinfo{author}{R.~Azuma}, \bibinfo{author}{E.~Barcikowski},
  \bibinfo{author}{J.~Belz}, \bibinfo{author}{D.~Bergman},
  \bibinfo{author}{S.~Blake}, \bibinfo{author}{R.~Cady},
  \bibinfo{author}{B.~Cheon}, \bibinfo{author}{J.~Chiba},
  \bibinfo{author}{M.~Chikawa}, \bibinfo{author}{E.~Cho},
  \bibinfo{author}{W.~Cho}, \bibinfo{author}{H.~Fujii},
  \bibinfo{author}{T.~Fujii}, \bibinfo{author}{T.~Fukuda},
  \bibinfo{author}{M.~Fukushima}, \bibinfo{author}{D.~Gorbunov},
  \bibinfo{author}{W.~Hanlon}, \bibinfo{author}{K.~Hayashi},
  \bibinfo{author}{Y.~Hayashi}, \bibinfo{author}{N.~Hayashida},
  \bibinfo{author}{K.~Hibino}, \bibinfo{author}{K.~Hiyama},
  \bibinfo{author}{K.~Honda}, \bibinfo{author}{T.~Iguchi},
  \bibinfo{author}{D.~Ikeda}, \bibinfo{author}{K.~Ikuta},
  \bibinfo{author}{N.~Inoue}, \bibinfo{author}{T.~Ishii},
  \bibinfo{author}{R.~Ishimori}, \bibinfo{author}{D.~Ivanov},
  \bibinfo{author}{S.~Iwamoto}, \bibinfo{author}{C.~Jui},
  \bibinfo{author}{K.~Kadota}, \bibinfo{author}{F.~Kakimoto},
  \bibinfo{author}{O.~Kalashev}, \bibinfo{author}{T.~Kanbe},
  \bibinfo{author}{K.~Kasahara}, \bibinfo{author}{H.~Kawai},
  \bibinfo{author}{S.~Kawakami}, \bibinfo{author}{S.~Kawana},
  \bibinfo{author}{E.~Kido}, \bibinfo{author}{H.~Kim},
  \bibinfo{author}{H.~Kim}, \bibinfo{author}{J.~Kim},
  \bibinfo{author}{K.~Kitamoto}, \bibinfo{author}{K.~Kobayashi},
  \bibinfo{author}{Y.~Kobayashi}, \bibinfo{author}{Y.~Kondo},
  \bibinfo{author}{K.~Kuramoto}, \bibinfo{author}{V.~Kuzmin},
  \bibinfo{author}{Y.~Kwon}, \bibinfo{author}{S.~Lim},
  \bibinfo{author}{S.~Machida}, \bibinfo{author}{K.~Martens},
  \bibinfo{author}{J.~Martineau}, \bibinfo{author}{T.~Matsuda},
  \bibinfo{author}{T.~Matsuura}, \bibinfo{author}{T.~Matsuyama},
  \bibinfo{author}{J.~Matthews}, \bibinfo{author}{I.~Myers},
  \bibinfo{author}{M.~Minamino}, \bibinfo{author}{K.~Miyata},
  \bibinfo{author}{H.~Miyauchi}, \bibinfo{author}{Y.~Murano},
  \bibinfo{author}{T.~Nakamura}, \bibinfo{author}{S.~Nam},
  \bibinfo{author}{T.~Nonaka}, \bibinfo{author}{S.~Ogio},
  \bibinfo{author}{M.~Ohnishi}, \bibinfo{author}{H.~Ohoka},
  \bibinfo{author}{K.~Oki}, \bibinfo{author}{D.~Oku},
  \bibinfo{author}{T.~Okuda}, \bibinfo{author}{A.~Oshima},
  \bibinfo{author}{S.~Ozawa}, \bibinfo{author}{I.~Park},
  \bibinfo{author}{M.~Pshirkov}, \bibinfo{author}{D.~Rodriguez},
  \bibinfo{author}{S.~Roh}, \bibinfo{author}{G.~Rubtsov},
  \bibinfo{author}{D.~Ryu}, \bibinfo{author}{H.~Sagawa},
  \bibinfo{author}{N.~Sakurai}, \bibinfo{author}{A.~Sampson},
  \bibinfo{author}{L.~Scott}, \bibinfo{author}{P.~Shah},
  \bibinfo{author}{F.~Shibata}, \bibinfo{author}{T.~Shibata},
  \bibinfo{author}{H.~Shimodaira}, \bibinfo{author}{B.~Shin},
  \bibinfo{author}{J.~Shin}, \bibinfo{author}{T.~Shirahama},
  \bibinfo{author}{J.~Smith}, \bibinfo{author}{P.~Sokolsky},
  \bibinfo{author}{T.~Sonley}, \bibinfo{author}{R.~Springer},
  \bibinfo{author}{B.~Stokes}, \bibinfo{author}{S.~Stratton},
  \bibinfo{author}{T.~Stroman}, \bibinfo{author}{S.~Suzuki},
  \bibinfo{author}{Y.~Takahashi}, \bibinfo{author}{M.~Takeda},
  \bibinfo{author}{A.~Taketa}, \bibinfo{author}{M.~Takita},
  \bibinfo{author}{Y.~Tameda}, \bibinfo{author}{H.~Tanaka},
  \bibinfo{author}{K.~Tanaka}, \bibinfo{author}{M.~Tanaka},
  \bibinfo{author}{S.~Thomas}, \bibinfo{author}{G.~Thomson},
  \bibinfo{author}{P.~Tinyakov}, \bibinfo{author}{I.~Tkachev},
  \bibinfo{author}{H.~Tokuno}, \bibinfo{author}{T.~Tomida},
  \bibinfo{author}{S.~Troitsky}, \bibinfo{author}{Y.~Tsunesada},
  \bibinfo{author}{K.~Tsutsumi}, \bibinfo{author}{Y.~Tsuyuguchi},
  \bibinfo{author}{Y.~Uchihori}, \bibinfo{author}{S.~Udo},
  \bibinfo{author}{H.~Ukai}, \bibinfo{author}{G.~Vasiloff},
  \bibinfo{author}{Y.~Wada}, \bibinfo{author}{T.~Wong},
  \bibinfo{author}{M.~Wood}, \bibinfo{author}{Y.~Yamakawa},
  \bibinfo{author}{H.~Yamaoka}, \bibinfo{author}{K.~Yamazaki},
  \bibinfo{author}{J.~Yang}, \bibinfo{author}{S.~Yoshida},
  \bibinfo{author}{H.~Yoshii}, \bibinfo{author}{R.~Zollinger},
  \bibinfo{author}{Z.~Zundel},
\newblock \bibinfo{title}{{The surface detector array of the Telescope Array
  experiment}},
\newblock \bibinfo{journal}{Nuclear Instruments and Methods in Physics Research
  Section A: Accelerators, Spectrometers, Detectors and Associated Equipment}
  \bibinfo{volume}{689} (\bibinfo{year}{2012}{\natexlab{b}})
  \bibinfo{pages}{87--97}.
\bibitem[{Fukushima et~al.(2006)Fukushima, Sokolsky, Abu-Zayyad, Aida, Allen,
  Anderson, Azuma, Barcikowski, Belz, Bergman, Blake, Cady, Cheon, Chiba,
  Chikawa, Cho, Cho, Fujii, Fujii, Fukuda, Gorbunov, Hanlon, Hayashi, Hayashi,
  Hayashida, Hibino, Hiyama, Honda, Iguchi, Ikeda, Ikuta, Inoue, Ishii,
  Ishimori, Ivanov, Iwamoto, Jui, Kadota, Kakimoto, Kalashev, Kanbe, Kasahara,
  Kawai, Kawakami, Kawana, Kido, Kim, Kim, Kim, Kitamoto, Kitamura, Kitamura,
  Kobayashi, Kobayashi, Kondo, Kuramoto, Kuzmin, Kwon, Lim, Machida, Martens,
  Martineau, Matsuda, Matsuura, Matsuyama, Matthews, Minamino, Miyata, Murano,
  Nagataki, Nakamura, Nam, Nonaka, Ogio, Ohnishi, Ohoka, Oki, Oku, Okuda,
  Oshima, Ozawa, Park, Pshirkov, Rodriguez, Roh, Rubtsov, Ryu, Sagawa, Sakurai,
  Sampson, Scott, Shah, Shibata, Shibata, Shimodaira, Shin, Shin, Shirahama,
  Smith, Sonley, Springer, Stokes, Stratton, Stroman, Suzuki, Takahashi,
  Takeda, Taketa, Takita, Tameda, Tanaka, Tanaka, Tanaka, Thomas, Thomson,
  Tinyakov, Tkachev, Tokuno, Tomida, Troitsky, Tsunesada, Tsutsumi, Tsuyuguchi,
  Uchihori, Udo, Ukai, Vasiloff, Wada, Wong, Wood, Yamakawa, Yamane, Yamaoka,
  Yamazaki, Yang, Yoneda, Yoshida, Yoshii, Zollinger, and Zundel}]{ICRR2006}
\bibinfo{author}{M.~Fukushima}, \bibinfo{author}{P.~Sokolsky},
  \bibinfo{author}{T.~Abu-Zayyad}, \bibinfo{author}{R.~Aida},
  \bibinfo{author}{M.~Allen}, \bibinfo{author}{R.~Anderson},
  \bibinfo{author}{R.~Azuma}, \bibinfo{author}{E.~Barcikowski},
  \bibinfo{author}{J.~W. Belz}, \bibinfo{author}{D.~R. Bergman},
  \bibinfo{author}{S.~A. Blake}, \bibinfo{author}{R.~Cady},
  \bibinfo{author}{B.~G. Cheon}, \bibinfo{author}{J.~Chiba},
  \bibinfo{author}{M.~Chikawa}, \bibinfo{author}{E.~J. Cho},
  \bibinfo{author}{W.~R. Cho}, \bibinfo{author}{H.~Fujii},
  \bibinfo{author}{T.~Fujii}, \bibinfo{author}{T.~Fukuda},
  \bibinfo{author}{D.~Gorbunov}, \bibinfo{author}{W.~Hanlon},
  \bibinfo{author}{K.~Hayashi}, \bibinfo{author}{Y.~Hayashi},
  \bibinfo{author}{N.~Hayashida}, \bibinfo{author}{K.~Hibino},
  \bibinfo{author}{K.~Hiyama}, \bibinfo{author}{K.~Honda},
  \bibinfo{author}{T.~Iguchi}, \bibinfo{author}{D.~Ikeda},
  \bibinfo{author}{K.~Ikuta}, \bibinfo{author}{N.~Inoue},
  \bibinfo{author}{T.~Ishii}, \bibinfo{author}{R.~Ishimori},
  \bibinfo{author}{D.~Ivanov}, \bibinfo{author}{S.~Iwamoto},
  \bibinfo{author}{C.~C.~H. Jui}, \bibinfo{author}{K.~Kadota},
  \bibinfo{author}{F.~Kakimoto}, \bibinfo{author}{O.~Kalashev},
  \bibinfo{author}{T.~Kanbe}, \bibinfo{author}{K.~Kasahara},
  \bibinfo{author}{H.~Kawai}, \bibinfo{author}{S.~Kawakami},
  \bibinfo{author}{S.~Kawana}, \bibinfo{author}{E.~Kido},
  \bibinfo{author}{H.~B. Kim}, \bibinfo{author}{H.~K. Kim},
  \bibinfo{author}{J.~H. Kim}, \bibinfo{author}{K.~Kitamoto},
  \bibinfo{author}{S.~Kitamura}, \bibinfo{author}{Y.~Kitamura},
  \bibinfo{author}{K.~Kobayashi}, \bibinfo{author}{Y.~Kobayashi},
  \bibinfo{author}{Y.~Kondo}, \bibinfo{author}{K.~Kuramoto},
  \bibinfo{author}{V.~Kuzmin}, \bibinfo{author}{Y.~J. Kwon},
  \bibinfo{author}{S.~I. Lim}, \bibinfo{author}{S.~Machida},
  \bibinfo{author}{K.~Martens}, \bibinfo{author}{J.~Martineau},
  \bibinfo{author}{T.~Matsuda}, \bibinfo{author}{T.~Matsuura},
  \bibinfo{author}{T.~Matsuyama}, \bibinfo{author}{J.~N. Matthews},
  \bibinfo{author}{M.~Minamino}, \bibinfo{author}{K.~Miyata},
  \bibinfo{author}{Y.~Murano}, \bibinfo{author}{S.~Nagataki},
  \bibinfo{author}{T.~Nakamura}, \bibinfo{author}{S.~W. Nam},
  \bibinfo{author}{T.~Nonaka}, \bibinfo{author}{S.~Ogio},
  \bibinfo{author}{M.~Ohnishi}, \bibinfo{author}{H.~Ohoka},
  \bibinfo{author}{K.~Oki}, \bibinfo{author}{D.~Oku},
  \bibinfo{author}{T.~Okuda}, \bibinfo{author}{A.~Oshima},
  \bibinfo{author}{S.~Ozawa}, \bibinfo{author}{I.~H. Park},
  \bibinfo{author}{M.~S. Pshirkov}, \bibinfo{author}{D.~C. Rodriguez},
  \bibinfo{author}{S.~Y. Roh}, \bibinfo{author}{G.~Rubtsov},
  \bibinfo{author}{D.~Ryu}, \bibinfo{author}{H.~Sagawa},
  \bibinfo{author}{N.~Sakurai}, \bibinfo{author}{A.~L. Sampson},
  \bibinfo{author}{L.~M. Scott}, \bibinfo{author}{P.~D. Shah},
  \bibinfo{author}{F.~Shibata}, \bibinfo{author}{T.~Shibata},
  \bibinfo{author}{H.~Shimodaira}, \bibinfo{author}{B.~K. Shin},
  \bibinfo{author}{J.~I. Shin}, \bibinfo{author}{T.~Shirahama},
  \bibinfo{author}{J.~D. Smith}, \bibinfo{author}{T.~J. Sonley},
  \bibinfo{author}{R.~W. Springer}, \bibinfo{author}{B.~T. Stokes},
  \bibinfo{author}{S.~R. Stratton}, \bibinfo{author}{T.~Stroman},
  \bibinfo{author}{S.~Suzuki}, \bibinfo{author}{Y.~Takahashi},
  \bibinfo{author}{M.~Takeda}, \bibinfo{author}{A.~Taketa},
  \bibinfo{author}{M.~Takita}, \bibinfo{author}{Y.~Tameda},
  \bibinfo{author}{H.~Tanaka}, \bibinfo{author}{K.~Tanaka},
  \bibinfo{author}{M.~Tanaka}, \bibinfo{author}{S.~B. Thomas},
  \bibinfo{author}{G.~B. Thomson}, \bibinfo{author}{P.~Tinyakov},
  \bibinfo{author}{I.~Tkachev}, \bibinfo{author}{H.~Tokuno},
  \bibinfo{author}{T.~Tomida}, \bibinfo{author}{S.~Troitsky},
  \bibinfo{author}{Y.~Tsunesada}, \bibinfo{author}{K.~Tsutsumi},
  \bibinfo{author}{Y.~Tsuyuguchi}, \bibinfo{author}{Y.~Uchihori},
  \bibinfo{author}{S.~Udo}, \bibinfo{author}{H.~Ukai},
  \bibinfo{author}{G.~Vasiloff}, \bibinfo{author}{Y.~Wada},
  \bibinfo{author}{T.~Wong}, \bibinfo{author}{M.~Wood},
  \bibinfo{author}{Y.~Yamakawa}, \bibinfo{author}{R.~Yamane},
  \bibinfo{author}{H.~Yamaoka}, \bibinfo{author}{K.~Yamazaki},
  \bibinfo{author}{J.~Yang}, \bibinfo{author}{Y.~Yoneda},
  \bibinfo{author}{S.~Yoshida}, \bibinfo{author}{H.~Yoshii},
  \bibinfo{author}{R.~Zollinger}, \bibinfo{author}{Z.~Zundel},
  \bibinfo{title}{{ICRR Annual Report}}, \bibinfo{type}{Technical Report}
  \bibinfo{number}{April 2006}, Institute for Cosmic Ray Research, University
  of Tokyo, \bibinfo{year}{2006}.
\bibitem[{Ivanov(2012)}]{Ivanov2012}
\bibinfo{author}{D.~Ivanov}, \bibinfo{title}{{Energy Spectrum Measured by the
  Telescope Array Surface Detector}}, Ph.D. thesis, Rutgers, the State
  University of New Jersey, \bibinfo{year}{2012}.
\bibitem[{Rodriguez(2011)}]{Rodriguez2011}
\bibinfo{author}{D.~C. Rodriguez}, \bibinfo{title}{{The Telescope Array Middle
  Drum Monocular Energy Spectrum and a Search for Coincident Showers Using High
  Resolution Fly's Eye HiRes-1 Monocular Data}}, Ph.D. thesis, University of
  Utah, \bibinfo{year}{2011}.
\bibitem[{Allen(2012)}]{Allen2012}
\bibinfo{author}{M.~G. Allen}, \bibinfo{title}{{Ultra High Energy Cosmic Ray
  Energy Spectrum and Composition Using Hybrid Analysis with Telescope Array}},
  Ph.D. thesis, University of Utah, \bibinfo{year}{2012}.
\bibitem[{Takeda et~al.(2002)Takeda, Sakaki, Honda, Chikawa, Fukushima,
  Hayashida, Inoue, Kadota, Kakimoto, Kamata, Kawaguchi, Kawakami, Kawasaki,
  Kawasumi, Mahrous, Mase, Mizobuchi, Morizane, Nagano, Ohoka, Osone, Sasaki,
  Sasano, Shimizu, Shinozaki, Teshima, Torii, Tsushima, Uchihori, Yamamoto,
  Yoshida, and Yoshii}]{Takeda2002}
\bibinfo{author}{M.~Takeda}, \bibinfo{author}{N.~Sakaki},
  \bibinfo{author}{K.~Honda}, \bibinfo{author}{M.~Chikawa},
  \bibinfo{author}{M.~Fukushima}, \bibinfo{author}{N.~Hayashida},
  \bibinfo{author}{N.~Inoue}, \bibinfo{author}{K.~Kadota},
  \bibinfo{author}{F.~Kakimoto}, \bibinfo{author}{K.~Kamata},
  \bibinfo{author}{S.~Kawaguchi}, \bibinfo{author}{S.~Kawakami},
  \bibinfo{author}{Y.~Kawasaki}, \bibinfo{author}{N.~Kawasumi},
  \bibinfo{author}{A.~M. Mahrous}, \bibinfo{author}{K.~Mase},
  \bibinfo{author}{S.~Mizobuchi}, \bibinfo{author}{Y.~Morizane},
  \bibinfo{author}{M.~Nagano}, \bibinfo{author}{H.~Ohoka},
  \bibinfo{author}{S.~Osone}, \bibinfo{author}{M.~Sasaki},
  \bibinfo{author}{M.~Sasano}, \bibinfo{author}{H.~M. Shimizu},
  \bibinfo{author}{K.~Shinozaki}, \bibinfo{author}{M.~Teshima},
  \bibinfo{author}{R.~Torii}, \bibinfo{author}{I.~Tsushima},
  \bibinfo{author}{Y.~Uchihori}, \bibinfo{author}{T.~Yamamoto},
  \bibinfo{author}{S.~Yoshida}, \bibinfo{author}{H.~Yoshii},
\newblock \bibinfo{title}{{Energy determination in the Akeno Giant Air Shower
  Array experiment}},
\newblock \bibinfo{journal}{Astroparticle Physics} \bibinfo{volume}{19}
  (\bibinfo{year}{2002}) \bibinfo{pages}{447--462}.
\bibitem[{Newton et~al.(2007)Newton, Knapp, and Watson}]{Newton2007}
\bibinfo{author}{D.~Newton}, \bibinfo{author}{J.~Knapp},
  \bibinfo{author}{a.~Watson},
\newblock \bibinfo{title}{{The optimum distance at which to determine the size
  of a giant air shower}},
\newblock \bibinfo{journal}{Astroparticle Physics} \bibinfo{volume}{26}
  (\bibinfo{year}{2007}) \bibinfo{pages}{414--419}.
\bibitem[{{Telescope Array
  Collaboration}(2014)}]{TelescopeArrayCollaboration2014}
\bibinfo{author}{{Telescope Array Collaboration}},
\newblock \bibinfo{title}{{CORSIKA Simulation of the Telescope Array Surface
  Detector}},
\newblock \bibinfo{journal}{Submitted to Astroparticle Physics}
  (\bibinfo{year}{2014}).
\bibitem[{Heck et~al.(1998)Heck, Knapp, Capdevielle, Schatz, and
  Thouw}]{Heck1998}
\bibinfo{author}{D.~Heck}, \bibinfo{author}{J.~Knapp}, \bibinfo{author}{J.~N.
  Capdevielle}, \bibinfo{author}{G.~Schatz}, \bibinfo{author}{T.~Thouw},
\newblock \bibinfo{title}{{CORSIKA: A Monte Carlo Code to Simulate Extensive
  Air Showers}},
\newblock \bibinfo{journal}{Forschungszentrum Karlsruhe} \bibinfo{volume}{FZKA
  6019} (\bibinfo{year}{1998}) \bibinfo{pages}{1--90}.
\bibitem[{Ostapchenko(2004)}]{Ostapchenko2004}
\bibinfo{author}{S.~Ostapchenko},
\newblock \bibinfo{title}{{QGSJET-II: towards reliable description of very high
  energy hadronic interactions}},
\newblock \bibinfo{journal}{Nuclear Physics B - Proceedings Supplements}
  \bibinfo{volume}{151} (\bibinfo{year}{2004}) \bibinfo{pages}{143--146}.
\bibitem[{Battistoni et~al.(2008)Battistoni, Garzelli, Gadioli, Muraro, Sala,
  Fass\`{o}, Ferrari, Roesler, Cerutti, Ranft, Pinsky, Empl, Pelliccioni, and
  Villari}]{Battistoni2006}
\bibinfo{author}{G.~Battistoni}, \bibinfo{author}{M.~V. Garzelli},
  \bibinfo{author}{E.~Gadioli}, \bibinfo{author}{S.~Muraro},
  \bibinfo{author}{P.~R. Sala}, \bibinfo{author}{A.~Fass\`{o}},
  \bibinfo{author}{A.~Ferrari}, \bibinfo{author}{S.~Roesler},
  \bibinfo{author}{F.~Cerutti}, \bibinfo{author}{J.~Ranft},
  \bibinfo{author}{L.~S. Pinsky}, \bibinfo{author}{A.~Empl},
  \bibinfo{author}{M.~Pelliccioni}, \bibinfo{author}{R.~Villari},
\newblock \bibinfo{title}{{The hadronic models for cosmic ray physics: the
  FLUKA code solutions}},
\newblock \bibinfo{journal}{Nuclear Physics B - Proceedings Supplements}
  \bibinfo{volume}{175-176} (\bibinfo{year}{2008}) \bibinfo{pages}{88--95}.
\bibitem[{Nelson et~al.(1985)Nelson, Hirayama, and Rogers}]{Nelson1985}
\bibinfo{author}{W.~R. Nelson}, \bibinfo{author}{H.~Hirayama},
  \bibinfo{author}{D.~W. Rogers}, \bibinfo{title}{{The EGS4 Code System}},
  \bibinfo{type}{Technical Report} \bibinfo{number}{SLAC-265}, Standford Linear
  Accelerator Center, \bibinfo{year}{1985}.
\bibitem[{Stokes et~al.(2012)Stokes, Cady, Ivanov, Matthews, and
  Thomson}]{Stokes2012}
\bibinfo{author}{B.~Stokes}, \bibinfo{author}{R.~Cady},
  \bibinfo{author}{D.~Ivanov}, \bibinfo{author}{J.~Matthews},
  \bibinfo{author}{G.~Thomson},
\newblock \bibinfo{title}{{Dethinning extensive air shower simulations}},
\newblock \bibinfo{journal}{Astroparticle Physics} \bibinfo{volume}{35}
  (\bibinfo{year}{2012}) \bibinfo{pages}{759--766}.
\bibitem[{Abbasi et~al.(2007)Abbasi, Abu-Zayyad, Allen, Amman, Archbold, Belov,
  Belz, {Ben Zvi}, Bergman, Blake, Brusova, Burt, Cannon, Cao, Connolly, Deng,
  Fedorova, Finley, Gray, Hanlon, Hoffman, Holzscheiter, Hughes,
  H\"{u}ntemeyer, Jones, Jui, Kim, Kirn, Loh, Maestas, Manago, Marek, Martens,
  Matthews, Matthews, Moore, O'Neill, Painter, Perera, Reil, Riehle, Roberts,
  Rodriguez, Sasaki, Schnetzer, Scott, Sinnis, Smith, Sokolsky, Song, Springer,
  Stokes, Thomas, Thomas, Thomson, Tupa, Westerhoff, Wiencke, Zhang, and
  Zech}]{Abbasi2007}
\bibinfo{author}{R.~U. Abbasi}, \bibinfo{author}{T.~Abu-Zayyad},
  \bibinfo{author}{M.~Allen}, \bibinfo{author}{J.~F. Amman},
  \bibinfo{author}{G.~Archbold}, \bibinfo{author}{K.~Belov},
  \bibinfo{author}{J.~W. Belz}, \bibinfo{author}{S.~Y. {Ben Zvi}},
  \bibinfo{author}{D.~R. Bergman}, \bibinfo{author}{S.~A. Blake},
  \bibinfo{author}{O.~A. Brusova}, \bibinfo{author}{G.~W. Burt},
  \bibinfo{author}{C.~Cannon}, \bibinfo{author}{Z.~Cao}, \bibinfo{author}{B.~C.
  Connolly}, \bibinfo{author}{W.~Deng}, \bibinfo{author}{Y.~Fedorova},
  \bibinfo{author}{C.~B. Finley}, \bibinfo{author}{R.~C. Gray},
  \bibinfo{author}{W.~F. Hanlon}, \bibinfo{author}{C.~M. Hoffman},
  \bibinfo{author}{M.~H. Holzscheiter}, \bibinfo{author}{G.~Hughes},
  \bibinfo{author}{P.~H\"{u}ntemeyer}, \bibinfo{author}{B.~F. Jones},
  \bibinfo{author}{C.~C.~H. Jui}, \bibinfo{author}{K.~Kim},
  \bibinfo{author}{M.~A. Kirn}, \bibinfo{author}{E.~C. Loh},
  \bibinfo{author}{M.~M. Maestas}, \bibinfo{author}{N.~Manago},
  \bibinfo{author}{L.~J. Marek}, \bibinfo{author}{K.~Martens},
  \bibinfo{author}{J.~A.~J. Matthews}, \bibinfo{author}{J.~N. Matthews},
  \bibinfo{author}{S.~A. Moore}, \bibinfo{author}{A.~O'Neill},
  \bibinfo{author}{C.~A. Painter}, \bibinfo{author}{L.~Perera},
  \bibinfo{author}{K.~Reil}, \bibinfo{author}{R.~Riehle},
  \bibinfo{author}{M.~Roberts}, \bibinfo{author}{D.~Rodriguez},
  \bibinfo{author}{N.~Sasaki}, \bibinfo{author}{S.~R. Schnetzer},
  \bibinfo{author}{L.~M. Scott}, \bibinfo{author}{G.~Sinnis},
  \bibinfo{author}{J.~D. Smith}, \bibinfo{author}{P.~Sokolsky},
  \bibinfo{author}{C.~Song}, \bibinfo{author}{R.~W. Springer},
  \bibinfo{author}{B.~T. Stokes}, \bibinfo{author}{S.~B. Thomas},
  \bibinfo{author}{J.~R. Thomas}, \bibinfo{author}{G.~B. Thomson},
  \bibinfo{author}{D.~Tupa}, \bibinfo{author}{S.~Westerhoff},
  \bibinfo{author}{L.~R. Wiencke}, \bibinfo{author}{X.~Zhang},
  \bibinfo{author}{A.~Zech},
\newblock \bibinfo{title}{{First observation of the Greisen-Zatsepin-Kuzmin
  suppression.}},
\newblock \bibinfo{journal}{Physical Review Letters} \bibinfo{volume}{100}
  (\bibinfo{year}{2007}) \bibinfo{pages}{4}.
\bibitem[{Abbasi et~al.(2014)Abbasi, Abe, Abu-Zayyad, Allen, Anderson, Azuma,
  Barcikowski, Belz, Bergman, Blake, Cady, Chae, Cheon, Chiba, Chikawa, Cho,
  Fujii, Fukushima, Goto, Hanlon, Hayashi, Hayashida, Hibino, Honda, Ikeda,
  Inoue, Ishii, Ishimori, Ito, Ivanov, Jui, Kadota, Kakimoto, Kalashev,
  Kasahara, Kawai, Kawakami, Kawana, Kawata, Kido, Kim, Kim, Kitamura,
  Kitamura, Kuzmin, Kwon, Lan, Lim, Lundquist, Machida, Martens, Matsuda,
  Matsuyama, Matthews, Minamino, Mukai, Myers, Nagasawa, Nagataki, Nakamura,
  Nonaka, Nozato, Ogio, Ogura, Ohnishi, Ohoka, Oki, Okuda, Ono, Oshima, Ozawa,
  Park, Pshirkov, Rodriguez, Rubtsov, Ryu, Sagawa, Sakurai, Sampson, Scott,
  Shah, Shibata, Shibata, Shimodaira, Shin, Shin, Smith, Sokolsky, Springer,
  Stokes, Stratton, Stroman, Suzawa, Takamura, Takeda, Takeishi, Taketa,
  Takita, Tameda, Tanaka, Tanaka, Tanaka, Thomas, Thomson, Tinyakov, Tkachev,
  Tokuno, Tomida, Troitsky, Tsunesada, Tsutsumi, Uchihori, Udo, Urban,
  Vasiloff, Wong, Yamane, Yamaoka, Yamazaki, Yang, Yashiro, Yoneda, Yoshida,
  Yoshiia, Zollinger, and Zundel}]{Abbasi2014}
\bibinfo{author}{R.~U. Abbasi}, \bibinfo{author}{M.~Abe},
  \bibinfo{author}{T.~Abu-Zayyad}, \bibinfo{author}{M.~Allen},
  \bibinfo{author}{R.~Anderson}, \bibinfo{author}{R.~Azuma},
  \bibinfo{author}{E.~Barcikowski}, \bibinfo{author}{J.~W. Belz},
  \bibinfo{author}{D.~R. Bergman}, \bibinfo{author}{S.~A. Blake},
  \bibinfo{author}{R.~Cady}, \bibinfo{author}{M.~J. Chae},
  \bibinfo{author}{B.~G. Cheon}, \bibinfo{author}{J.~Chiba},
  \bibinfo{author}{M.~Chikawa}, \bibinfo{author}{W.~R. Cho},
  \bibinfo{author}{T.~Fujii}, \bibinfo{author}{M.~Fukushima},
  \bibinfo{author}{T.~Goto}, \bibinfo{author}{W.~Hanlon},
  \bibinfo{author}{Y.~Hayashi}, \bibinfo{author}{N.~Hayashida},
  \bibinfo{author}{K.~Hibino}, \bibinfo{author}{K.~Honda},
  \bibinfo{author}{D.~Ikeda}, \bibinfo{author}{N.~Inoue},
  \bibinfo{author}{T.~Ishii}, \bibinfo{author}{R.~Ishimori},
  \bibinfo{author}{H.~Ito}, \bibinfo{author}{D.~Ivanov},
  \bibinfo{author}{C.~C.~H. Jui}, \bibinfo{author}{K.~Kadota},
  \bibinfo{author}{F.~Kakimoto}, \bibinfo{author}{O.~Kalashev},
  \bibinfo{author}{K.~Kasahara}, \bibinfo{author}{H.~Kawai},
  \bibinfo{author}{S.~Kawakami}, \bibinfo{author}{S.~Kawana},
  \bibinfo{author}{K.~Kawata}, \bibinfo{author}{E.~Kido},
  \bibinfo{author}{H.~B. Kim}, \bibinfo{author}{J.~H. Kim},
  \bibinfo{author}{S.~Kitamura}, \bibinfo{author}{Y.~Kitamura},
  \bibinfo{author}{V.~Kuzmin}, \bibinfo{author}{Y.~J. Kwon},
  \bibinfo{author}{J.~Lan}, \bibinfo{author}{S.~I. Lim}, \bibinfo{author}{J.~P.
  Lundquist}, \bibinfo{author}{K.~Machida}, \bibinfo{author}{K.~Martens},
  \bibinfo{author}{T.~Matsuda}, \bibinfo{author}{T.~Matsuyama},
  \bibinfo{author}{J.~N. Matthews}, \bibinfo{author}{M.~Minamino},
  \bibinfo{author}{Y.~Mukai}, \bibinfo{author}{I.~Myers},
  \bibinfo{author}{K.~Nagasawa}, \bibinfo{author}{S.~Nagataki},
  \bibinfo{author}{T.~Nakamura}, \bibinfo{author}{T.~Nonaka},
  \bibinfo{author}{A.~Nozato}, \bibinfo{author}{S.~Ogio},
  \bibinfo{author}{J.~Ogura}, \bibinfo{author}{M.~Ohnishi},
  \bibinfo{author}{H.~Ohoka}, \bibinfo{author}{K.~Oki},
  \bibinfo{author}{T.~Okuda}, \bibinfo{author}{M.~Ono},
  \bibinfo{author}{A.~Oshima}, \bibinfo{author}{S.~Ozawa},
  \bibinfo{author}{I.~H. Park}, \bibinfo{author}{M.~S. Pshirkov},
  \bibinfo{author}{D.~C. Rodriguez}, \bibinfo{author}{G.~Rubtsov},
  \bibinfo{author}{D.~Ryu}, \bibinfo{author}{H.~Sagawa},
  \bibinfo{author}{N.~Sakurai}, \bibinfo{author}{A.~L. Sampson},
  \bibinfo{author}{L.~M. Scott}, \bibinfo{author}{P.~D. Shah},
  \bibinfo{author}{F.~Shibata}, \bibinfo{author}{T.~Shibata},
  \bibinfo{author}{H.~Shimodaira}, \bibinfo{author}{B.~K. Shin},
  \bibinfo{author}{H.~S. Shin}, \bibinfo{author}{J.~D. Smith},
  \bibinfo{author}{P.~Sokolsky}, \bibinfo{author}{R.~W. Springer},
  \bibinfo{author}{B.~T. Stokes}, \bibinfo{author}{S.~R. Stratton},
  \bibinfo{author}{T.~Stroman}, \bibinfo{author}{T.~Suzawa},
  \bibinfo{author}{M.~Takamura}, \bibinfo{author}{M.~Takeda},
  \bibinfo{author}{R.~Takeishi}, \bibinfo{author}{A.~Taketa},
  \bibinfo{author}{M.~Takita}, \bibinfo{author}{Y.~Tameda},
  \bibinfo{author}{H.~Tanaka}, \bibinfo{author}{K.~Tanaka},
  \bibinfo{author}{M.~Tanaka}, \bibinfo{author}{S.~B. Thomas},
  \bibinfo{author}{G.~B. Thomson}, \bibinfo{author}{P.~Tinyakov},
  \bibinfo{author}{I.~Tkachev}, \bibinfo{author}{H.~Tokuno},
  \bibinfo{author}{T.~Tomida}, \bibinfo{author}{S.~Troitsky},
  \bibinfo{author}{Y.~Tsunesada}, \bibinfo{author}{K.~Tsutsumi},
  \bibinfo{author}{Y.~Uchihori}, \bibinfo{author}{S.~Udo},
  \bibinfo{author}{F.~Urban}, \bibinfo{author}{G.~Vasiloff},
  \bibinfo{author}{T.~Wong}, \bibinfo{author}{R.~Yamane},
  \bibinfo{author}{H.~Yamaoka}, \bibinfo{author}{K.~Yamazaki},
  \bibinfo{author}{J.~Yang}, \bibinfo{author}{K.~Yashiro},
  \bibinfo{author}{Y.~Yoneda}, \bibinfo{author}{S.~Yoshida},
  \bibinfo{author}{H.~Yoshiia}, \bibinfo{author}{R.~Zollinger},
  \bibinfo{author}{Z.~Zundel},
\newblock \bibinfo{title}{{Study of Ultra-High Energy Cosmic Ray Composition
  Using Telescope Array's Middle Drum Detector and Surface Array in Hybrid
  Mode}}  (\bibinfo{year}{2014}) \bibinfo{pages}{22}.
\bibitem[{Jui(2012)}]{Jui2012}
\bibinfo{author}{C.~C.~H. Jui},
\newblock \bibinfo{title}{{Cosmic Ray in the Northern Hemisphere: Results from
  the Telescope Array Experiment}},
\newblock \bibinfo{journal}{Journal of Physics: Conference Series}
  \bibinfo{volume}{404} (\bibinfo{year}{2012}) \bibinfo{pages}{012037}.

\end{thebibliography}
